\documentclass{aa}  
\usepackage{graphics,graphicx}
\usepackage{xcolor}
\usepackage{enumitem}
\usepackage{float}
\usepackage{caption}
\usepackage{subcaption}

\usepackage{txfonts}
\usepackage{longtable}
\usepackage[unicode=true,pdfusetitle,
 bookmarks=true,bookmarksnumbered=false,bookmarksopen=false,
 breaklinks=false,pdfborder={0 0 0},pdfborderstyle={},backref=false,colorlinks=true]
 {hyperref}
 
\hypersetup{
 urlcolor=blue,citecolor=blue}

\begin{document}  
   \title{Chemical enrichment of ICM within the A3266 cluster I: radial profiles.}  
   \author{E. Gatuzz\inst{1},   
           J. Sanders\inst{1}, 
           A. Liu\inst{1,2}, 
           A. Fabian\inst{3}, 
           C. Pinto\inst{4}, 
           H. Russell\inst{5},   
           D. Eckert\inst{6}, \\ 
           S. Walker\inst{7}
           J. ZuHone\inst{8}   \and
           R. Mohapatra\inst{9}
          } 
 
   \institute{Max-Planck-Institut f\"ur extraterrestrische Physik, Gie{\ss}enbachstra{\ss}e 1, 85748 Garching, Germany\\
              \email{egatuzz@mpe.mpg.de}
         \and
             Institute for Frontiers in Astronomy and Astrophysics, Beijing Normal University, Beijing 102206, China
         \and
             Institute of Astronomy, Madingley Road, Cambridge CB3 0HA, UK
         \and
              INAF - IASF Palermo, Via U. La Malfa 153, I-90146 Palermo, Italy 
          \and
             School of Physics \& Astronomy, University of Nottingham, University Park, Nottingham NG7 2RD, UK
          \and
             Department of Astronomy, University of Geneva, Ch. d\rq Ecogia 16, CH-1290 Versoix, Switzerland
          \and
             Department of Physics and Astronomy, University of Alabama in Huntsville, Huntsville, AL 35899, USA
          \and
             Harvard-Smithsonian Center for Astrophysics, 60 Garden Street, Cambridge, MA, 02138, USA         
          \and
             Department of Astrophysical Sciences, Princeton University, NJ 08544, USA                       
             }
 
   \date{Received XXX; accepted YYY}  
  \abstract  
{We present a detailed study of the elemental abundances distribution of the intracluster medium (ICM) within the A3266 cluster using {\it XMM-Newton} observations. 
This analysis uses EPIC-pn data, including a new energy scale calibration, which allows us to measure velocities with uncertainties down to $\Delta v \sim 80$ km/s, and MOS observations.
We measured radial O, Mg, Si, S, Ar, Ca, and Fe profiles.
This is the first study of elemental abundances beyond Fe using X-ray observations within the A3266 cluster. 
The abundance profiles display discontinuities similar to those obtained for the temperature. 
We modeled the X/Fe ratio profiles with a linear combination of type~Ia supernovae (SNIa) and core-collapse supernovae (SNcc) models.
We found that the SNIa ratio over the total cluster enrichment tends to be uniform, with a $42\pm 5\%$ contribution.
Such a trend supports an early ICM enrichment scenario, with most metals produced before clustering.  

} 
 
\keywords{X-rays: galaxies: clusters - Galaxies: clusters: intracluster medium - Galaxies: clusters: individual: A3266 - }
\titlerunning{Radial profiles of the A3266 Cluster}
\authorrunning{Gatuzz et al.}
\maketitle

\section{Introduction}\label{sec_in}   
The intracluster medium (ICM) hosts most of the baryonic mass in a cluster of galaxies, thus providing crucial information on the origin and distribution of chemical elements during the evolution of the Universe.
Fe-peak elements (Cr, Mn, Fe, Ni) mainly originate from type-Ia supernovas (SNIa), and light $\alpha$-elements (O, Ne, Mg) are mainly produced in core-collapse supernovae (SNcc). 
Intermediate-mass elements (e.g., Si, S, Ar, and Ca) are synthesized by both SNIa and SNcc \citep[e.g.,][and references therein]{nom13}.
In this way, the ICM abundance distribution is sensitive to the number of SNIa and SNcc that contribute to chemical enrichment, the initial mass function (IMF) of the stars that explode as SNcc, the initial metallicity of the progenitors, the SNIa explosion mechanism and the time scale over which the supernova products are expelled  \citep[see][for a review]{wer08}.
However, there is an intense debate regarding the SNIa explosion mechanisms, with the two main theoretical models proposed including pure deflagration \citep{ple04,ple07,kas07,fin14,lon14} and delayed denotation \citep{hoe96,iwa99,gam05,rop12}.
Therefore, accurate ICM abundance measurements will help to distinguish between both scenarios.

X-ray spectroscopy of the ICM provides a powerful tool to constrain the abundance distribution of these elements from their emission lines \citep{gat23b,gat23d,gat23e}.
A central Fe abundance excess has been found in many cool-core clusters \citep[e.g.][]{deg01,chu03,pan15,mer17,liu19,liu20}.
Towards the outskirts, a flat and azimuthally uniform Fe distribution has been observed \citep{mat11,wer13,sim15,urb17}.
Furthermore, {\it Hitomi} observations of the Perseus cluster show abundance ratios entirely consistent with Solar \citep{hit18,sim19}.
These results point out the importance of the contribution from both SNIa (with sub-Chandrasekhar mass) and near SNcc to the chemical enrichment of the ICM. 

The A3266 galaxy cluster constitutes an excellent target to study the chemical enrichment of the ICM within a merging system.
\citet{hen02} measured a nonmonotonically decreasing radial Fe profile using {\it Chandra}.
A Fe abundance map was created by \citet{sau05} with {\it XMM-Newton} data.
They found that the Fe abundance map indicates that metals are inhomogeneously distributed across the cluster due to the merging features of the system. However, they did not measure individual elemental abundances.
A3266 was a calibration target of the {\it SRG/eROSITA} X-ray telescope \citep{pre21} on board the {\it Spectrum Roentgen Gamma} (SRG) observatory.
In their data analysis, \citet{san22a} identified three different subclusters merging with the main body of the cluster. 
These systems appear associated with higher-metallicity gas.

\citet{gat24a} studied the velocity structure within the A3266 galaxy cluster using {\it XMM-Newton} observations.
They observed that the hot gas moves with a redshifted systemic velocity relative to the cluster core across all fields of view.
Here, we present a detailed study of the elemental abundance distribution within the A3266 galaxy cluster.
The outline of this paper is as follows.
We describe the data reduction process in Section~\ref{sec_dat}.
In Section \ref{sec_fits}, we explain the fitting procedure. 
A discussion of the results is shown in Section~\ref{sec_dis}.
Finally, Section~\ref{sec_con} includes the conclusions and summary.  
Throughout this paper we assume the distance of A3266 to be $z=0.0594$ \citep{deh17} and a concordance $\Lambda$CDM cosmology with $\Omega_m = 0.3$, $\Omega_\Lambda = 0.7$, and $H_{0} = 70 \textrm{ km s}^{-1}\ \textrm{Mpc}^{-1}$.

\section{Data reduction}\label{sec_dat} 
We used the same observations analyzed in \citet{gat24a} and followed the reduction process shown in \citet[][c]{san20,gat22a}. 
The {\it XMM-Newton} European Photon Imaging Camera \citep[EPIC,][]{str01} spectra were reduced with the Science Analysis System (SAS\footnote{\url{https://www.cosmos.esa.int/web/xmm-newton/sas}}, version 21.0.0). 
We decided to include spectra from both MOS and pn instruments in our analysis, operating in full frame mode and extended full frame mode, respectively.
We reduce the pn, MOS~1, and MOS~2 data using the SAS {\tt epchain} and {\tt emchain} tools.
We filtered the data using FLAG==0 to avoid bad pixels or regions close to CCD edges.
We used the single, double, triple, and quadruple events in MOS (PATTERN $\leq$ 12), while we used only single-pixel events for the pn instrument (PATTERN==0).
We also identified and discarded those EPIC MOS chips in the anomalous state using the {\tt emanom} task.
We created the EPIC-pn event files, including the new energy calibration scale developed in \citet{san20,gat22a}, to obtain velocity measurements down to 150~km/s at Fe-K by using the instrumental X-ray background lines as references for the absolute energy scale.
Point-like sources were identified and removed from the subsequent analysis by using the SAS {\tt edetect\_chain} tool, with a likelihood parameter {\tt det\_ml} $> 10$

\begin{figure}    
\centering
\includegraphics[width=0.49\textwidth]{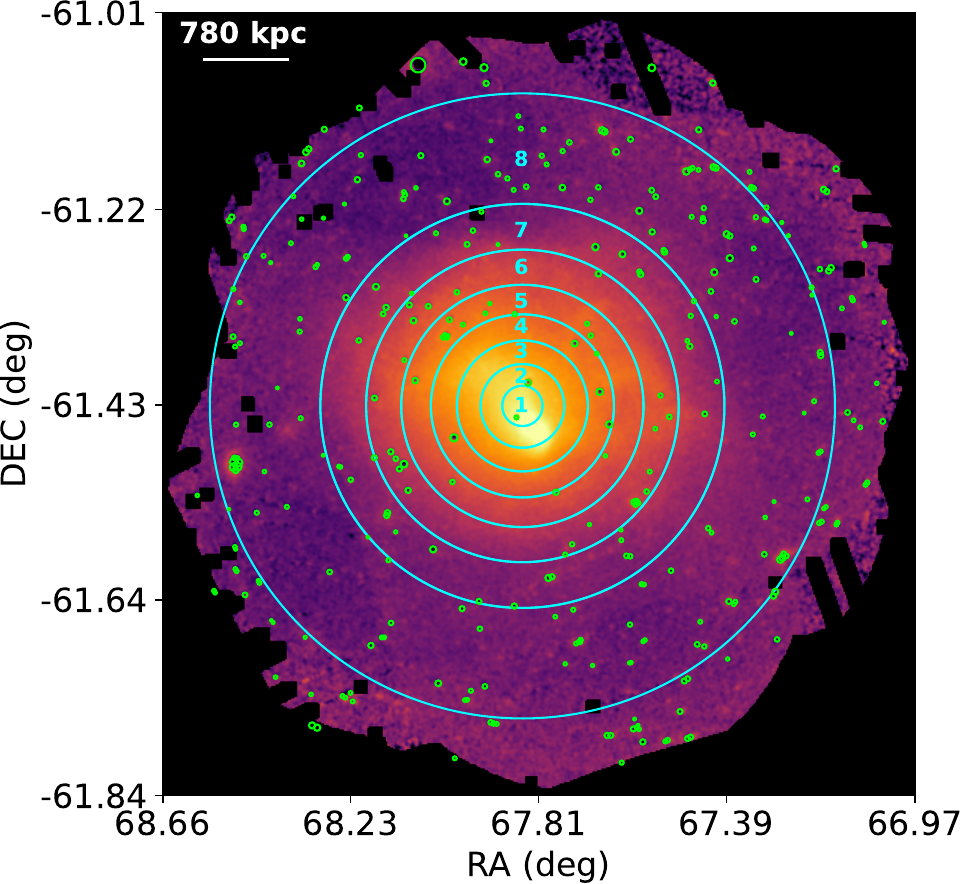} 
\caption{Exposure vignetting corrected and background-subtracted X-ray image of the A3266 cluster in the 0.5-9.25~keV energy range obtained for the EPIC-pn instrument. 
Cyan rings correspond to the analyzed regions, while green circles correspond to point sources that were excluded from the analysis.
} \label{fig_regions} 
\end{figure}

For the pn camera, we filtered bad time intervals from flares with a 1.0 cts/s rate threshold to be consistent with the \citet{gat24a} procedure.
We build the Good Time Intervales for the MOS cameras by applying the $2\sigma$ clipping technique \citep{mer15}, which we now explain in detail.
First, we extract the count-rate histogram from the light curves in the $10-12$~keV energy range.
After binning them in 100 s intervals, we fitted them with a Gaussian function, and we reject all time bins for which the number of counts lies outside the interval $\mu\pm 2\sigma$, where $\mu$ is the fitted average of the distribution.
Then, we repeat the same procedure for 10 s binned histograms in the $0.3-10$~keV energy range to remove soft background flares.
We iterate through this process and manually inspect the light curves until all flares are removed.
To study the distribution of chemical elements in the ICM, we analyzed non-overlapping rings whose radii increase as the square root with distance from the A3266 center. 
Figure~\ref{fig_regions} shows the exposure vignetting corrected and a background-subtracted X-ray image of the A3266 cluster in the 0.5-9.25~keV energy range obtained for the EPIC-pn instrument.
The figure also included the annular rings corresponding to the spectra extraction regions.
The radius of the annular regions increases up to $\sim$1125~kpc.
The obtained spectra have more than $5\times 10^{5}$ counts in the analyzed energy range.   

We extracted the EPIC-pn spectra following the procedure described by \citet{san20}. 
The spectra were extracted using the {\tt evselect} task, covering channels 0-20479. 
The response matrix was generated with the {\tt rmfgen} task, employing a fine energy grid with a resolution of 0.3~eV. 
Auxiliary Response Files (ARFs) were created using the {\tt arfgen} task, with the {\tt setbackscale} and {\tt extendedsource} parameters set to {\tt yes}.
The MOS spectra were extracted using the {\tt evselect} task, covering channels 0-11999. 
The response files were also extracted using the {\tt rmfgen} task, spanning the 0.1-12~keV energy range with a total of 1190 energy bins. 
The ARFs were generated using the {\tt arfgen} task, with the parameter {\tt extendedsource} set to {\tt yes}.
For the EPIC spectra, we applied binning using the standard factor of five PI channels (i.e., 5~eV bins).

\section{Spectral fitting}\label{sec_fits}  
We combine the spectra from different observations for each spatial region shown in Figure~\ref{fig_regions}.
The new EPIC-pn energy calibration scale cannot be applied to lower energies; therefore, we load the data twice to fit separately, but simultaneously, the soft ($0.5-4.0$~keV) and hard ($4.0-10$~keV) energy bands.
The previous work in \citet{gat24a} only included the hard band.
The MOS spectra were fitted in the $0.5-10$~keV energy range.
We fit the spectra with the {\it xspec} spectral fitting package (version 12.14.1\footnote{\url{https://heasarc.gsfc.nasa.gov/xanadu/xspec/}}). 
Abundances are given relative to \citet{lod09}.
We rebin the data to have at least 1 count per channel and we assumed {\tt cash} statistics \citep{cas79}.
Errors are quoted at 1$\sigma$ confidence level unless otherwise stated. 

We assume that the ICM is in collisional ionization equilibrium (CIE), and we use a single {\tt vapec} thermal model to fit the spectra.
{\tt vapec} is a variant of the widely used {\tt apec} model, including variable abundances most abundant elements. 
The free parameters include the temperature ($T_{e}$), elemental abundances (O, Mg, Si, Ar, S, Ca, Fe), redshift, and normalization.
To account for the Galactic absorption, we included a {\tt tbabs} component \citep{wil00}, and fixed the column density to $2.26\times 10^{20}$ cm$^{-2}$ \citep{HI416}.
As discussed in \citet{gat23b,gat23d,gat23e}, we set the Al, and Ni abundances to the Fe abundance, as their measurements may not be reliable due to instrumental background lines.
We also link the Ne abundance to Fe, as the Fe L-lines are located in the same spectral region, possibly leading to confusion between the lines.
Other abundances with $Z\geq6$ are also fixed to the Fe value.

Fitting the EPIC spectra over an extensive energy range can lead to slight under- or overestimation of the continuum emission in specific bands due to imperfections in the EPIC instrument calibration \citep[see][]{mer15, mer16}.
These discrepancies can, in turn, affect the derived elemental abundances, as the equivalent width (EW) of emission lines depends on the ratio between the line and the continuum flux.
To correct these effects, we employed a local-fit method that refines the derived abundances by fitting narrower energy ranges centered around the strongest emission lines of each element.
\citet{mer15} found that systematics in individual abundance measurements are reduced to $<15\%$ when applying this local fit method, thus more accurately determining line-to-continuum ratios. 
In this approach, we first performed a global spectra fit using the energy range of $0.5-10$~keV for both EPIC-pn and MOS.
We re-fitted the spectra within local energy bands to refine the abundance measurements. 
We chose to focus on the strongest K-shell emission lines of each element (except Ne, whose strongest lines are unresolved in the EPIC spectra). 
The energy ranges for the local fits are $0.50-0.80$ keV (for O), $1.20-1.65$ keV (for Mg), $1.65-2.26$ keV (for Si), $2.27-3$ keV (for S and Ar), $3.5-4.5$ keV (for Ca) and $6.25-10.$ keV (for Fe).
During these local fits, the global best-fit temperature was fixed to preserve the global thermal structure of the plasma. 
In contrast, the local normalization (emission measure) and the abundance of the element of interest were left free. 
This conservative approach accounts for systematic uncertainties related to EPIC cross-calibration issues and ensures robust abundance measurements. 
Additionally, we verified that the local-fit abundances were consistent within 2$\sigma$ with the global-fit results for most elements, confirming that the local corrections did not introduce significant biases.
All abundance measurements presented in this work are corrected hereafter using the local-fit method.

The background model consists of two main components: the instrumental background and the astrophysical background.
The astrophysical background consists of the unresolved population of point sources, modeled by a power-law with $\Gamma=1.45$ \citep{san20}, the Local Hot Bubble (LHB) emission, modeled by one unabsorbed thermal plasma model ({\tt apec}) and the Galactic halo emission \citep[GH, e.g.][]{yos09}, which consist of one absorbed thermal plasma model ({\tt apec}). 
For the thermal models, we consider the temperatures obtained for A3266 by \citet{san22} in their analysis with the {\it SRG/eROSITA} X-ray telescope \citep{pre21} on board of the {\it Spectrum Roentgen Gamma} (SRG) observatory.
In particular, we fixed the temperatures to $0.133$~keV for the LHB and $0.241$~kev for the GH.
The abundance of both components was kept proto-solar.

\begin{table*}
\caption{\label{tab_circular_fits}A3266 cluster best-fit parameters for annular regions.}
\centering
\begin{tabular}{ccccccccc}
\\
Parameter &Region 1&Region 2&Region 3&Region 4&Region 5&Region 6&Region 7&Region 8\\
\hline
\multicolumn{9}{c}{ICM component}\\
$kT$& $9.68_{-0.17}^{+0.14}$ & $7.61\pm 0.12$ & $7.01_{-0.12}^{+0.16}$ & $6.70 \pm 0.08$ & $7.03\pm 0.14$ & $7.65 \pm 0.12$ & $7.73\pm 0.13$ & $6.62 \pm 0.07$  \\
O& $1.16\pm 0.49$ & $0.43\pm 0.30$ & $0.44\pm 0.27$ & $0.42\pm 0.25$ & $0.49\pm 0.29$ & $0.84\pm 0.35$ & $0.93\pm 0.37$ & $0.49 \pm 0.19$  \\
Mg& $0.67\pm 0.30$ & $0.35 \pm 0.21$ & $0.25 \pm 0.19$ & $0.28 \pm 0.13$ & $0.46 \pm 0.19$ & $0.73 \pm 0.22$ & $0.49\pm 0.23$ & $0.41 \pm 0.13$  \\
Si& $0.32 \pm 0.18$ & $0.33 \pm 0.13$ & $0.38 \pm 0.12$ & $0.46\pm 0.11$ & $0.45 \pm 0.12$ & $0.54\pm 0.14$ & $0.39 \pm 0.14$ & $0.54 \pm 0.08$  \\
S& $0.52\pm 0.26$ & $0.27 \pm 0.18$ & $0.35 \pm 0.17$ & $0.63 \pm 0.15$ & $0.64 \pm 0.17$ & $0.49 \pm 0.19$ & $0.54 \pm 0.2$ & $0.30 \pm 0.11$  \\
Ar& $2.64 \pm 0.62$ & $1.72 \pm 0.45$ & $1.73\pm 0.41$ & $1.37 \pm 0.36$ & $1.16\pm 0.40$ & $1.91\pm 0.47$ & $1.87 \pm 0.48$ & $1.35 \pm 0.28$  \\
Ca& $0.80\pm 0.65$ & $0.44_{-0.39}^{+0.34}$ & $0.41\pm 0.37$ & $0.56 \pm 0.40$ & $0.64\pm 0.46$ & $1.09 \pm 0.51$ & $0.69\pm 0.53$ & $0.68 \pm 0.31$  \\
Fe& $0.57 \pm 0.06$ & $0.61\pm 0.06$ & $0.58 \pm 0.05$ & $0.53 \pm 0.04$ & $0.52\pm 0.05$ & $0.6 \pm 0.05$ & $0.63\pm 0.06$ & $0.55\pm 0.04$  \\
$z$& $6.15_{-0.13}^{+0.05}$ & $6.15 \pm 0.04$ & $6.15 \pm 0.04$ & $6.01_{-0.05}^{+0.02}$ & $6.16 \pm 0.03$ & $6.00 \pm 0.03$ & $5.84_{-0.05}^{+0.02}$ & $5.98\pm 0.03$  \\
$norm$& $3.18_{-0.54}^{+0.33}$ & $6.12_{-0.36}^{+0.46}$ & $7.04_{-0.33}^{+0.50}$ & $5.82_{-0.31}^{+0.39}$ & $4.88_{-0.41}^{+0.36}$ & $4.26_{-0.36}^{+0.40}$ & $4.05_{-0.34}^{+0.47}$ & $9.36_{-0.30}^{+0.53}$  \\
\hline
\multicolumn{9}{c}{Astrophysical background}\\
LHB$_{norm}$& $0.46\pm 0.02$ & $0.55\pm 0.02$ & $0.57\pm 0.02$ & $0.59\pm 0.02$ & $0.60\pm 0.02$ & $0.62\pm 0.02$ & $0.66\pm 0.02$ & $0.83\pm 0.02$  \\
GH$_{norm}$& $0.14\pm 0.02$ & $0.17\pm 0.02$ & $0.19\pm 0.02$ & $0.18\pm 0.02$ & $0.19\pm 0.02$ & $0.17\pm 0.02$ & $0.21\pm 0.02$ & $0.30\pm 0.02$  \\ 
{\tt pow}$_{norm}$& $1.67\pm 0.10$ & $2.19\pm 0.10$ & $2.31\pm 0.11$ & $2.78\pm 0.08$ & $2.86\pm 0.10$ & $2.64\pm 0.10$ & $2.69\pm 0.10$ & $2.63\pm 0.10$  \\
\hline
\multicolumn{9}{c}{Best-fit statistic}\\
cstat/d.o.f.& $6933/5701$ & $6951/5701$ & $6941/5701$ & $7084/5701$ & $7208/5701$ & $7257/5701$ & $7311/5701$ & $7232/5701$  \\
\hline
\multicolumn{8}{l}{$z$ in units of $\times 10^{-2}$, {\tt vapec}$_{norm}$ in units of $\times 10^{-3}$. Astrophysical background in units of $\times 10^{-3}$.}\\ 
\end{tabular}
\end{table*}
 
The instrumental background consists of a broken power-law (not folded by the effective area) to model the hard particle background \citep{mer15}, a power-law (not folded by the effective area) to model the soft-proton background \citep{erd21} and several instrumental profiles. 
This approach follows the methodology that we have already applied in our previous analysis of the Virgo \citep{gat23b}, Centaurus \citep{gat23c} and Ophiuchus \citep{gat23e} galaxy clusters.
For the EPIC-pn spectra, we included the Al K$\alpha$ (1.49~keV), Ni K$\alpha$ (7.84~keV), Cu K$\alpha$ (8.04~keV), Zn K$\alpha$ (8.64~keV) and Cu K$\beta$ (8.90~keV)  instrumental background lines.
We included the Al K$\alpha$ (1.49~keV) and Si K$\alpha$ (1.75~keV) for the MOS spectra.
The redshift of the background lines was tied together and allowed to vary.
The normalizations of the lines were free to vary.
We fixed the soft-proton background for the EPIC-pn instrument to $\Gamma=0.136$ for consistency with \citep{san20}.
Following \citet{erd21}, we fixed the breaking point of the broken power-law at $3.0$~keV.
The lower limit of the photon index ($\Gamma_{1}$) was left free to vary within 0.1-1.4, and the upper limit of the photon index ($\Gamma_{2}$) was allowed to reach up to 2.5.
We initially obtained best-fitting values for the hard particle and soft-proton background for the entire field-of-view (i.e., a circular region with a radius equal to the outer border of the ring eight shown in Figure~\ref{fig_regions}). 
Then, we kept only the normalization as a free parameter in the fits of the individual rings.
The normalizations of the different background model components vary across the regions analyzed during the global fit.
During the local fits, performed to estimate the abundances better, the background model normalizations are fixed to the value obtained in the global fit.

\begin{figure}    
\centering
\includegraphics[width=0.49\textwidth]{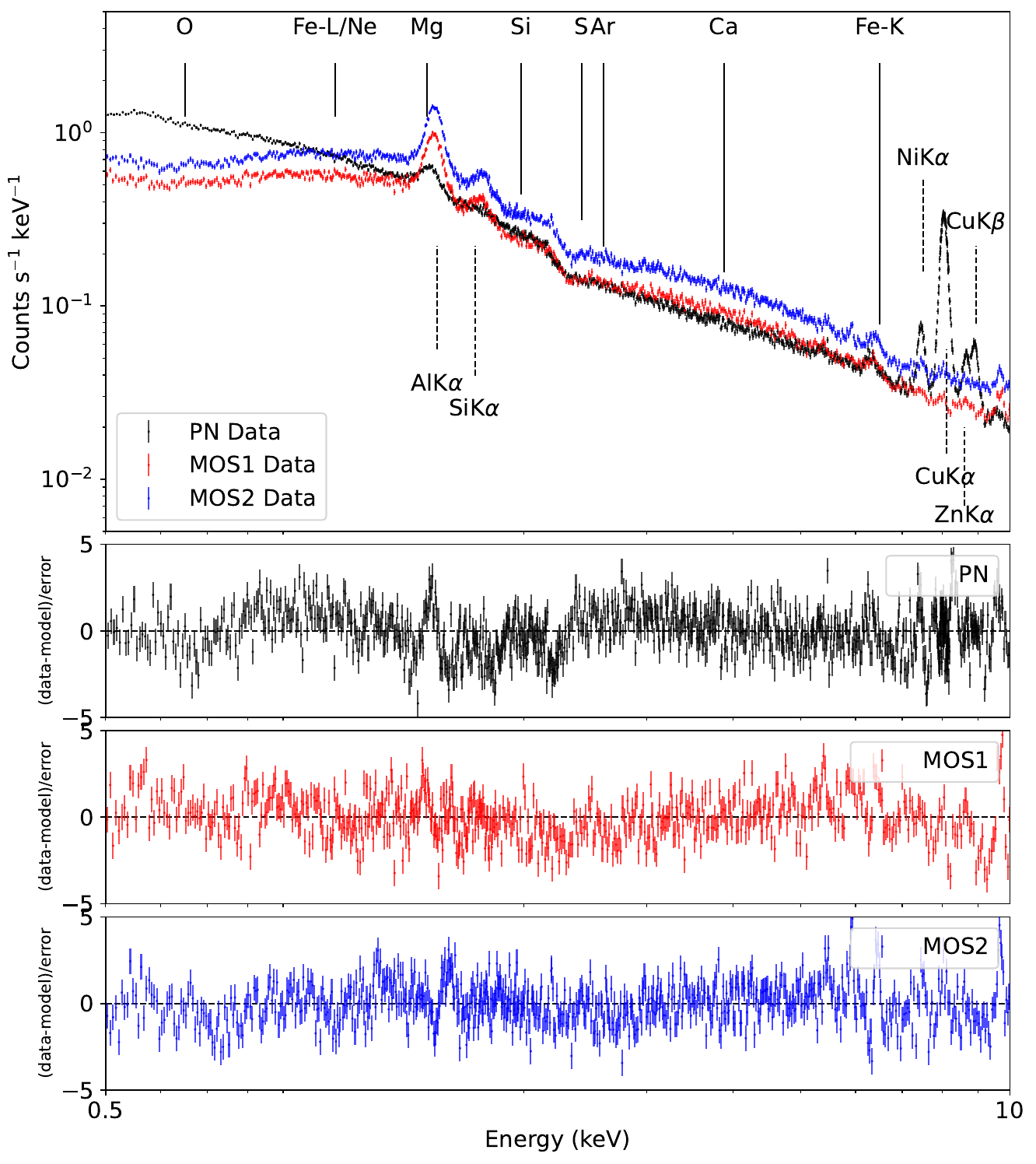} 
\caption{
Example spectrum and best-fit model obtained from the global fit for Region 2. 
The spectrum has been rebinned for illustrative purposes. EPIC-pn (black points), MOS1 (red points), and MOS2 (blue points) have been included. 
The line contribution from instrumental background (vertical dashed lines) and ICM emission (vertical solid lines) are indicated. 
The lower panel shows the residuals of the fit for each instrument.
}\label{fig_spectra_example} 
\end{figure}

\begin{figure}    
\centering
\includegraphics[width=0.48\textwidth]{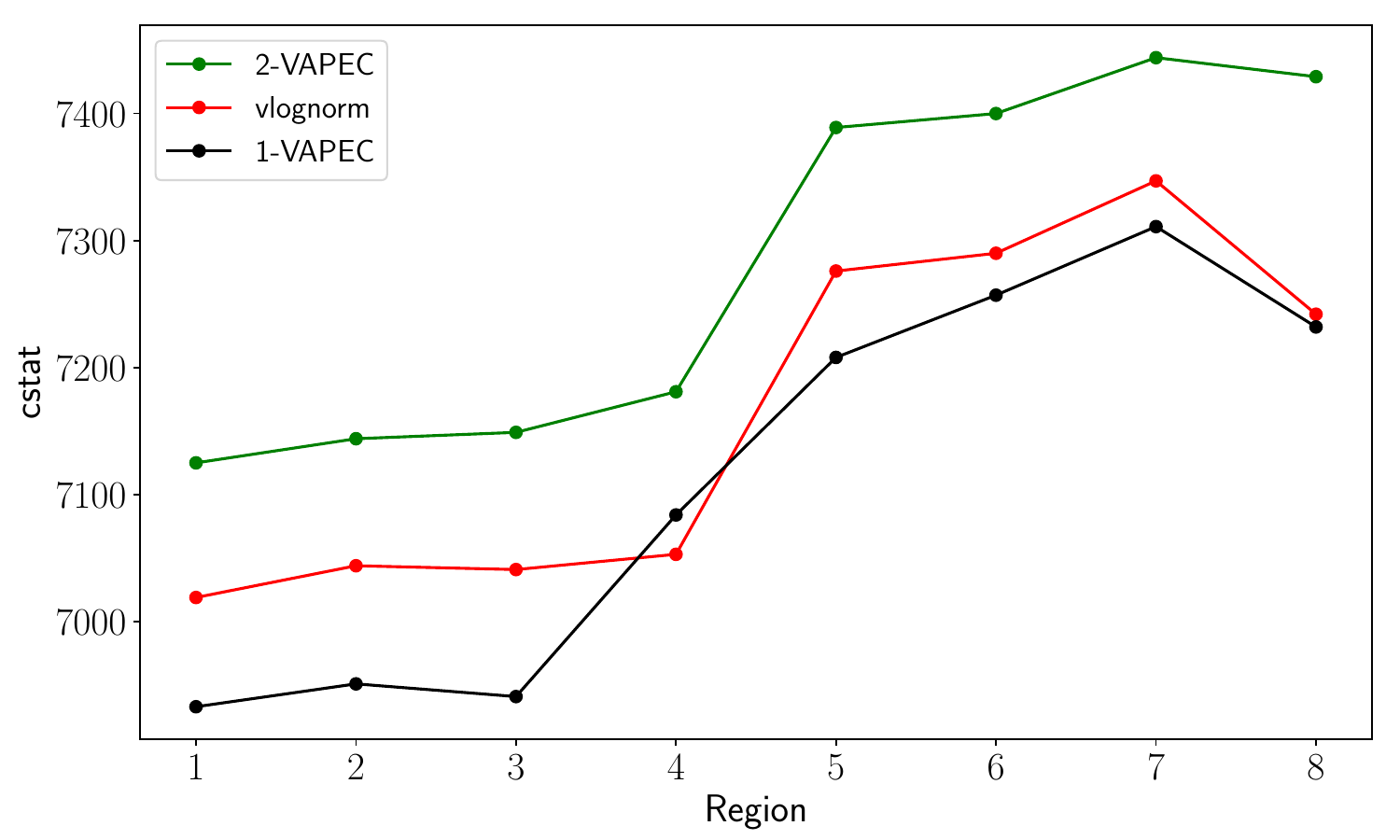} 
\caption{ 
Best-fitting Cash statistic obtained with the 1-{\tt apec}, 2-{\tt apec}, and {\tt lognorm} models.  
}\label{fig_cstat} 
\end{figure}

Figure~\ref{fig_spectra_example} shows an example spectrum and residuals from this analysis (i.e., for region 2). 
The solid line indicates the best-fitting {\tt apec} model. 
Vertical dashed lines indicate the instrumental Al K$\alpha$, Si K$\alpha$, Ni K$\alpha$, Cu K$\alpha$,$\beta$, and Zn K$\alpha$ background lines included in the model. 
Vertical solid lines indicate the contribution of O, Mg, Si, S, Ar, Ca, and Fe to the line emission.
Table~\ref{tab_circular_fits} shows the best-fit parameters obtained for each ring.
We noted that the background component increases as we move towards the cluster outskirts.
Previous studies have found such variation and reflect the spectral decomposition and degeneracy with cluster emission \citep{urb11,wal19}.
In the inner regions where the cluster emission is bright, the contribution of the soft X-ray background component LHB and GB is relatively minor.
In contrast, where the cluster surface brightness decreases at larger radii, the relative contribution of the LHB and GB components becomes more prominent, and their normalizations are better constrained.
Furthermore, the GH emission exhibits some degree of spatial variation on large scales due to differences in Galactic column density and the presence of local structures in the hot gas distribution  \citep{yos09,hen13}.
To test the significance of this variation, we performed fits with fixed LHB and GH normalizations (determined from the complete region) and found that while the overall fit statistic worsened slightly, the cluster parameters remained consistent within uncertainties. 
This suggests that while there is an apparent increase in the background normalizations, the effect does not strongly bias our primary cluster measurements.  
We also tested additional models for the ICM component, including a two {\tt apec} thermal components model and a {\tt lognorm} model \citep{gat23d}.
Figure~\ref{fig_cstat} compares the best-fitting statistic obtained for each model.
The numbering is from the innermost to the outermost.
However, the best-fit c-stat value remained very similar for the different models. 
Therefore, we decided to analyze the A3266 cluster using the single-temperature model.

To assess the impact of background modeling uncertainties on our measured abundances, we performed a test in which we varied the normalizations of the instrumental and astrophysical background components by $\pm 30\%$. 
We found that increasing the background by 30$\%$ resulted in a significantly worse fit (e.g., the cstat increase from $cstat=6933$ to $cstat=$13695 in the innermost region). 
Such a large degradation in fit quality suggests that a 30$\%$ increase is not a physically reasonable scenario for our data.  
Given that our best-fit background models have uncertainties of less than 10$\%$, we performed a more realistic test by varying the background components within $\pm 10\%$. In this case, the measured abundances remained consistent within statistical uncertainties, indicating that our abundance measurements are robust against reasonable variations in background modeling.   
This test also suggests that the abundances of elements with weak emission lines, such as O, Si, S, Ar, and Ca, are not significantly biased by continuum fitting effects. 
The consistency of the abundance measurements under realistic background variations reinforces the reliability of our spectral fitting approach.

\section{Results and discussion}\label{sec_dis} 
In this Section we discuss the best-fit results obtained. 

\subsection{Temperature profile}\label{spec_maps}  
\begin{figure}    
\centering
\includegraphics[width=0.48\textwidth]{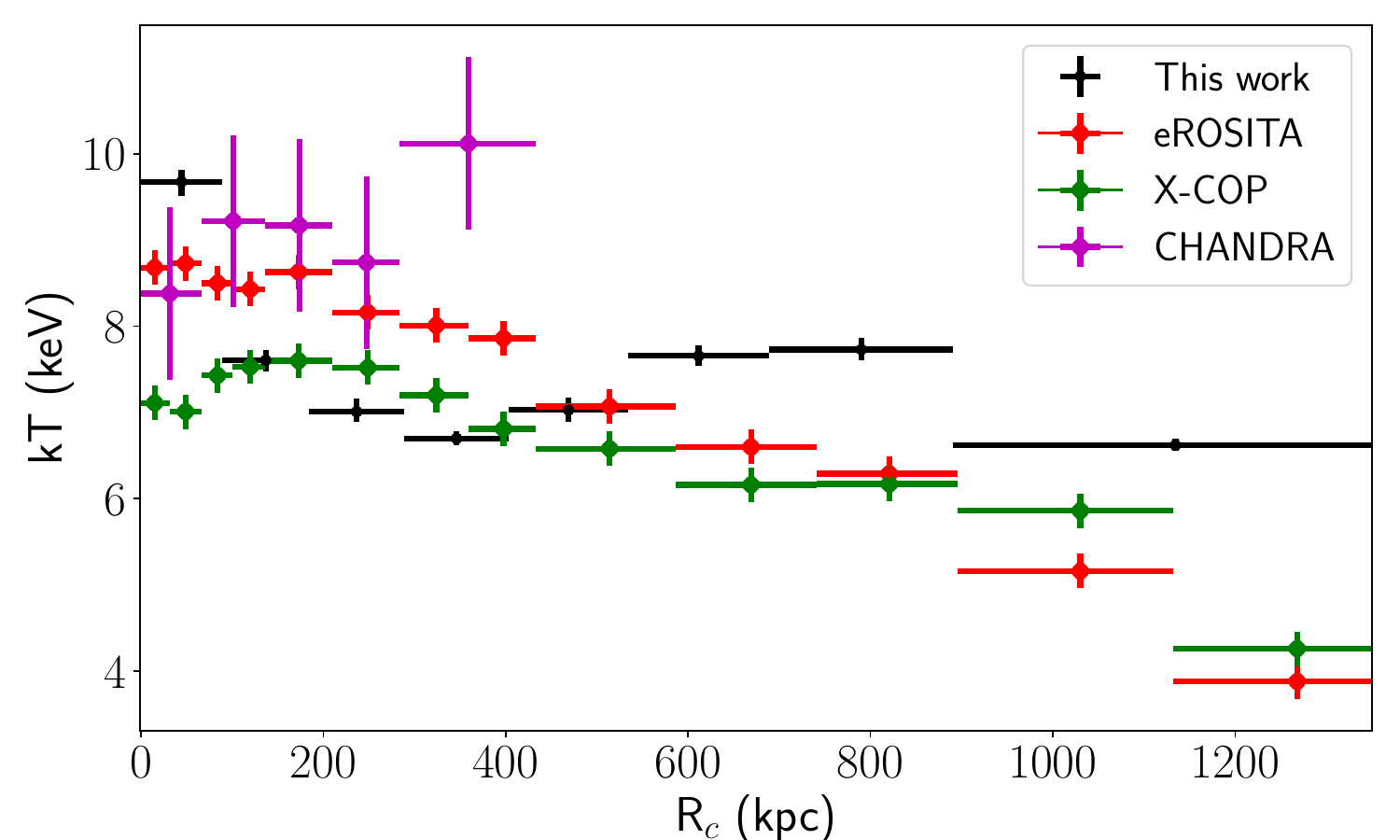} 
\caption{ 
\emph{Top panel:} Temperature profile obtained from the global fit.  
The Figure also includes the temperature profiles obtained with {\it eROSITA} \citep{san22a}, X-COP \citep{ghi19} and {\it Chandra} \citep{hen02}.
} \label{fig_kt_sigma} 
\end{figure}

Figure~\ref{fig_kt_sigma} shows the temperature values obtained from the best fit per region. 
The temperatures are generally similar to those obtained by \citet{gat24a}, except for regions 1 and 8, where temperatures here obtained are larger and lower, respectively.
We identified evident discontinuities at $\sim 350$~kpc and $\sim 800$~kpc, which can also be observed in the surface brightness map computed by \citep{gat24a}.  
The temperature in the $ 600-800 $ ~kpc range is constant.
Such discontinuities could be associated with subgroups within the system, as identified by \citet{deh17}. 

Figure~\ref{fig_kt_sigma} also includes the temperature profiles obtained with {\it eROSITA} \citep[][red points]{san22a}, X-COP \citep[][green points]{ghi19} and {\it Chandra} \citep[][magenta points]{hen02}
\citet{sau05} also found large temperatures ($kT>12$) for the inner region. 
Temperatures are also similar to those obtained by \citet{san22a}.
The high temperature near the cluster core could be due to Active Galactic Nuclei (AGN) activity.
A similar trend of temperature increasing and then decreasing for distances $>800$~kpc was found in {\it Chandra} data \citep{hen02}.

The discrepancy between the temperatures obtained for large distances from different telescopes may be due to several aspects, including calibration differences, as discussed in \citet{san22}.
The differences between our measured temperature profile for A3266 and the results presented in the X-COP study can be attributed to several factors. 
First, our study benefits from approximately eight times the exposure time used in X-COP. 
Furthermore, we use $N_{\rm H}=2.26\times 10^{20}$ cm$^{-2}$, while X-COP adopts $N_{\rm H}=1.62\times 10^{20}$ cm$^{-2}$, leading to systematic differences in the derived temperatures, particularly at lower energies where absorption is more significant. 
Additionally, our analysis includes all spectra in the $0.5-10$~keV energy range, whereas X-COP excludes specific energy bands ($1.2-1.9$~keV for MOS, $1.2-1.7$~keV and $7.0-9.2$~keV for pn) to mitigate instrumental background effects. 
The exclusion of these bands in X-COP likely impacts the temperature measurements, especially in regions dominated by background. 
Differences in atomic data also contribute, as we employ the latest APEC version with updated atomic data, while X-COP used an earlier version. 
Background modeling differences and the treatment of element abundances also play a role; our analysis allows individual element abundances to vary as free parameters, unlike X-COP, which scales all abundances to Fe. 
This distinction affects the coupling between temperature and metallicity in spectral fitting and can impact the resulting temperature profile. 
Notably, when we fix $N_{\rm H}$ to the X-COP value and adopt a simplified abundance model with all abundances equal to Fe, the temperature profile at large radii decreases and aligns more closely with the X-COP results. 
These considerations explain the observed discrepancies with X-COP results.

\subsection{Velocity profile}\label{spec_maps} 
\begin{figure}    
\centering  
\includegraphics[width=0.48\textwidth]{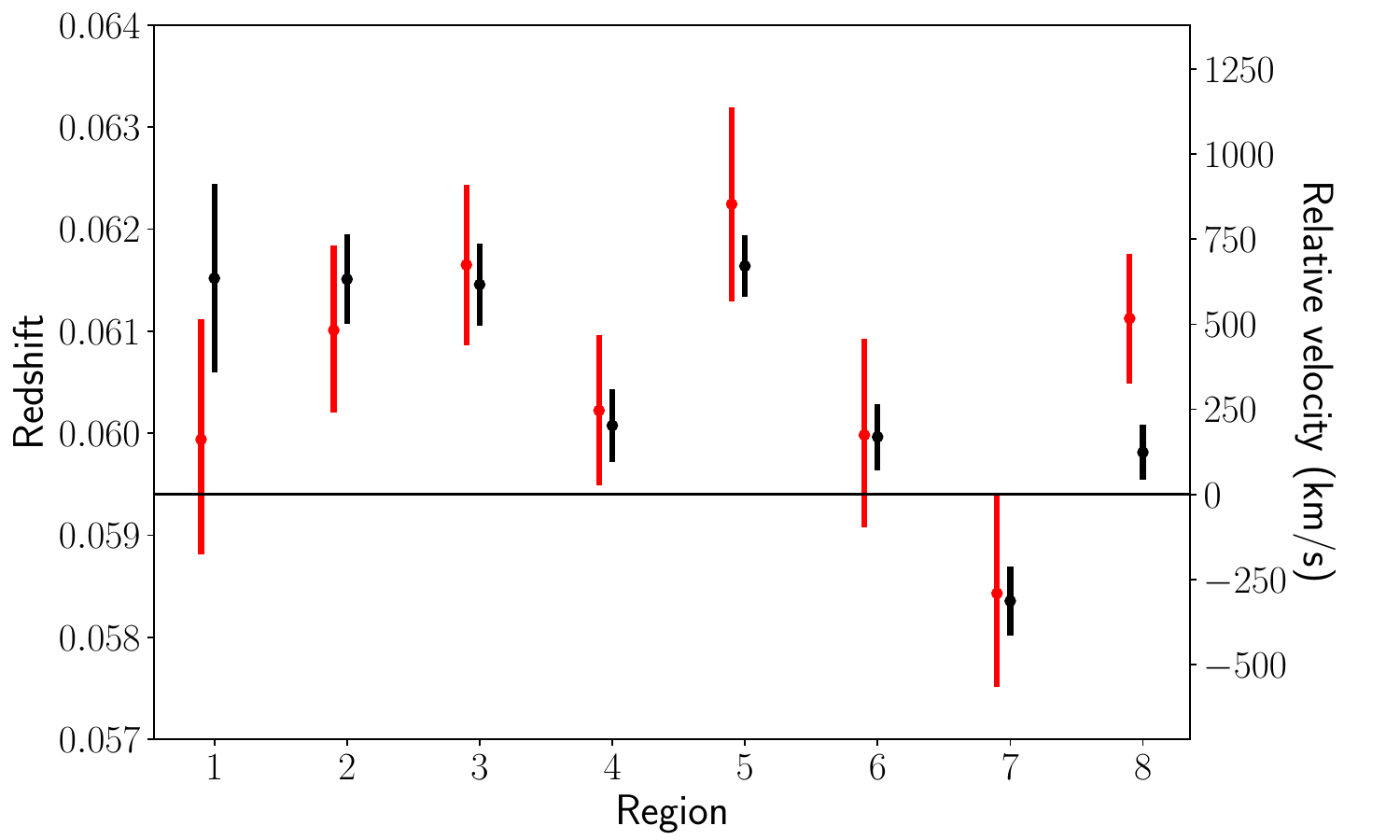}   
\caption{
A comparison between the velocities obtained for each region (black points) and those obtained by \citet[][ red points]{gat24a}. 
The horizontal line indicates the A3266 cluster redshift ($z=0.0594$). 
}\label{fig_vel} 
\end{figure}

The velocities obtained for each region are shown in Figure~\ref{fig_vel}.
As in our previous series of analyses of the ICM in galaxy clusters, these velocities correspond to the redshift obtained with the {\tt vapec} model with respect to the source redshift ($z=0.0594$).
This redshift corresponds to the mean redshift of the cluster core and was obtained from spectroscopic redshifts of 1300 sources within the cluster by \citet{deh17}. 
We have obtained velocities with uncertainties down to $\Delta v\sim 80$ km/s (region 8). 
The largest blueshift/redshift obtained, with respect to the A3266 systemic redshift, correspond to $635\pm 276$~km/s (region 1) and $-313\pm 102$~km/s (region 7).
Figure~\ref{fig_vel} also includes the velocities obtained by \citep{gat24a}, which agree with the current results, even though we have included the soft-energy band.
Moreover, including the MOS data leads to a better constraint on the velocity measurement. 
The significant discrepancy in the outermost region compared to \citep{gat24a} likely arises from a combination of factors: (1) higher astrophysical background in the cluster outskirts (see Table~\ref{tab_circular_fits}), which may affect the accuracy of Fe~K$\alpha$ centroid measurements, especially in low signal-to-background regions; (2) projection effects and dynamical complexity in the cluster outskirts; and (3) the inclusion of EPIC-MOS spectra, which reduces velocity uncertainties overall but may amplify sensitivity to background modeling in regions with low signal.
Notably, the velocities agree elsewhere, suggesting that the discrepancy in the outermost region is driven by the challenging observational conditions there.

\subsection{Abundance profiles}\label{circle_rings}
\begin{figure}    
\centering
\includegraphics[width=0.48\textwidth]{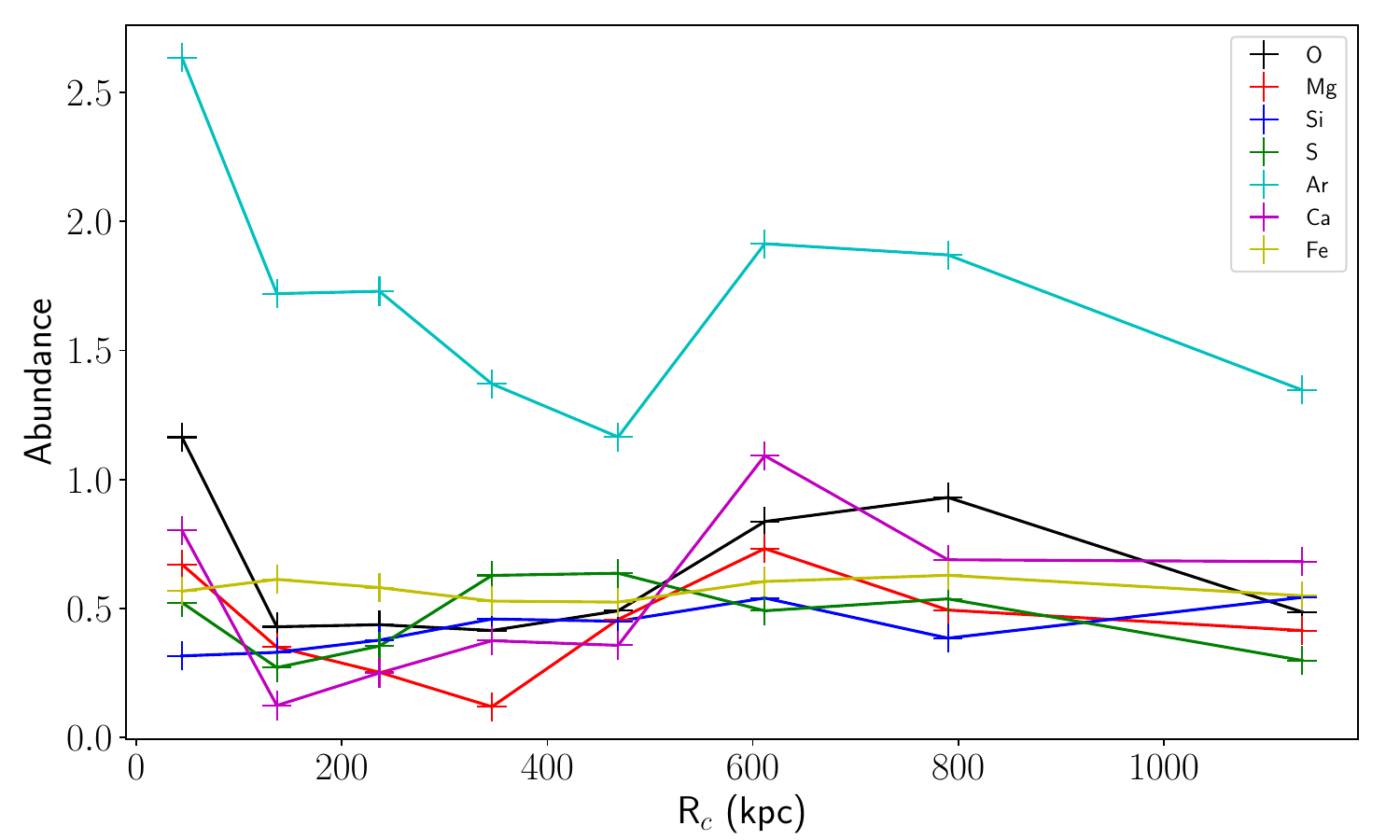}  
\caption{
A3266 radial abundances distribution for each element. Uncertainties are not included for illustrative purposes.
} \label{fig_abund_together} 
\end{figure}

\begin{figure*}    
\centering
\includegraphics[width=0.33\textwidth]{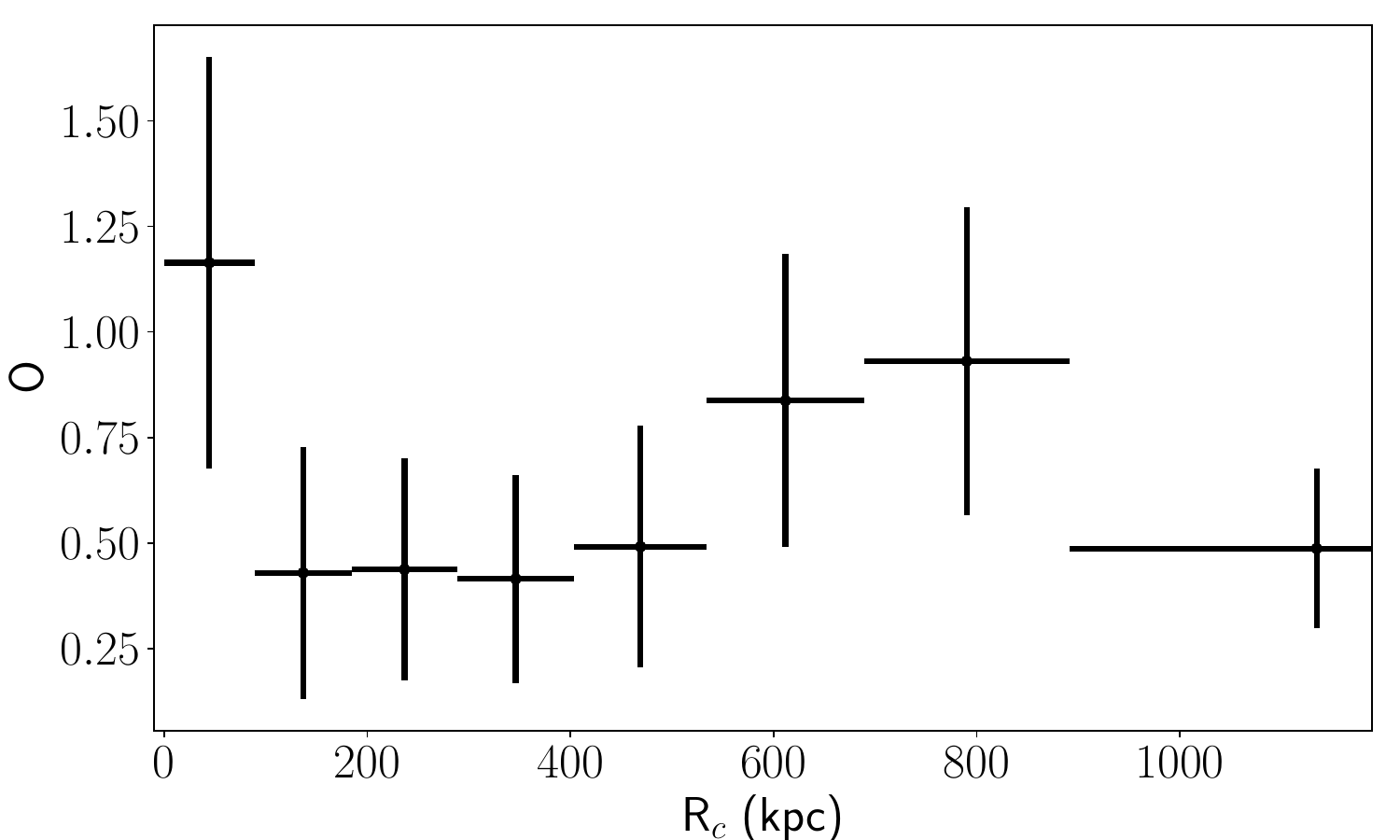} 
\includegraphics[width=0.33\textwidth]{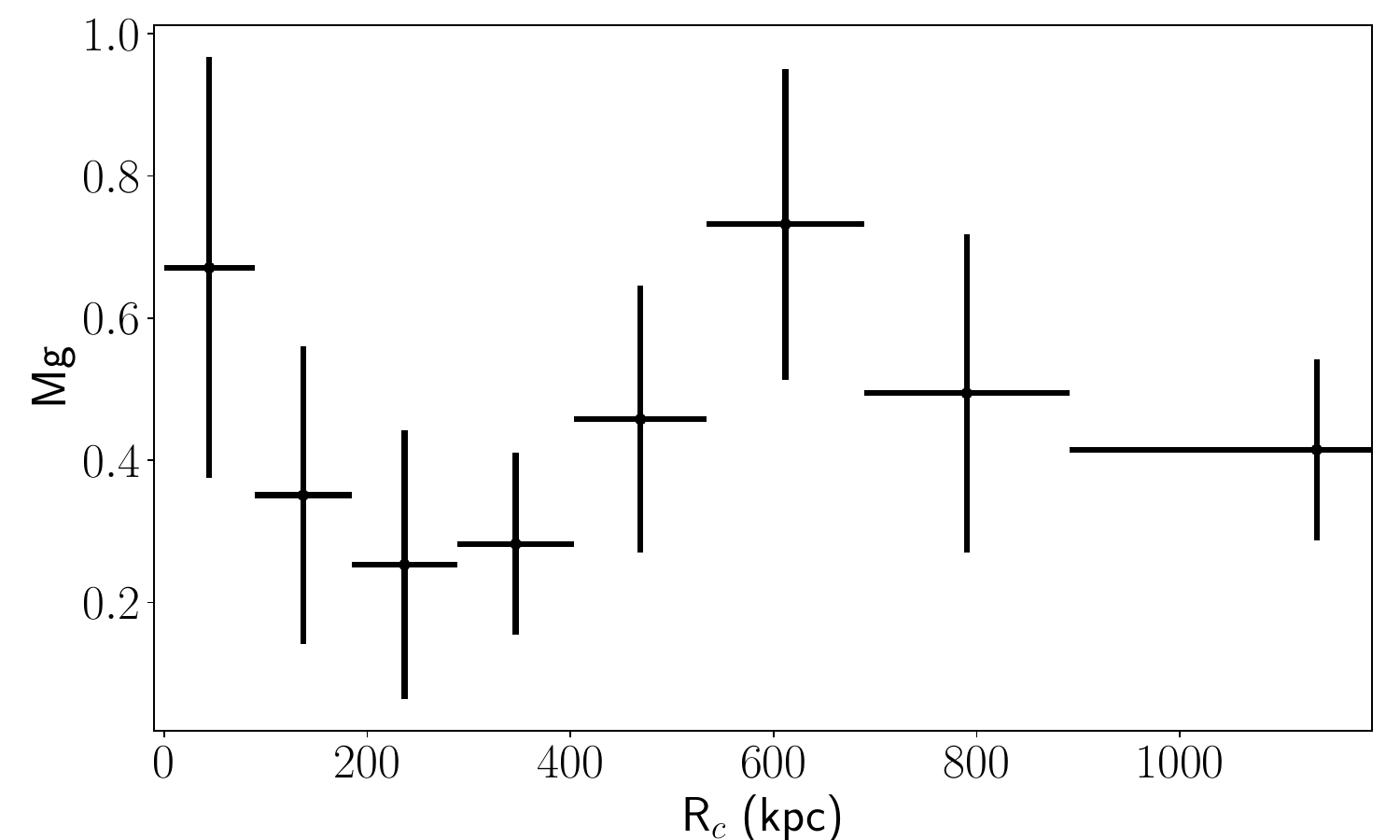}
\includegraphics[width=0.33\textwidth]{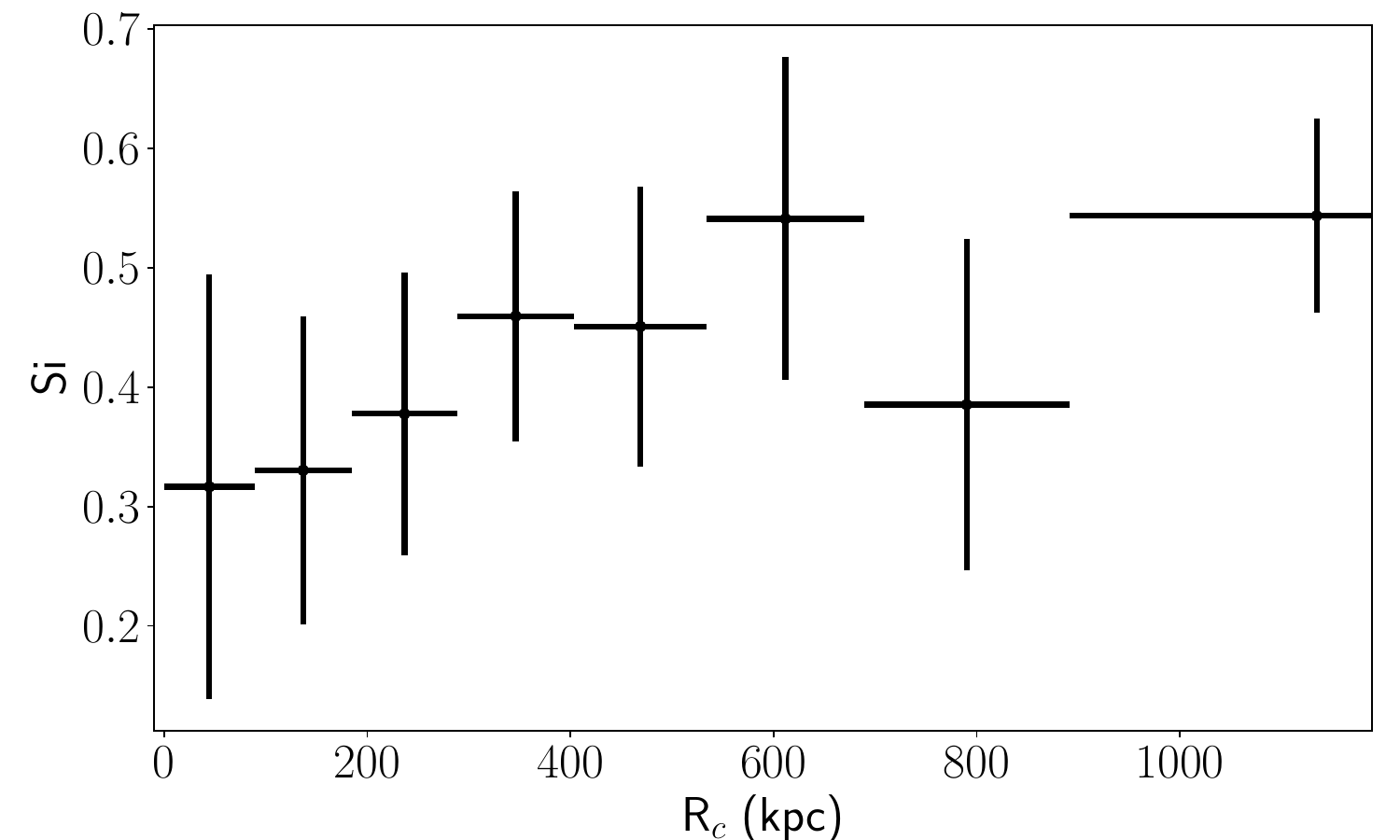}\\
\includegraphics[width=0.33\textwidth]{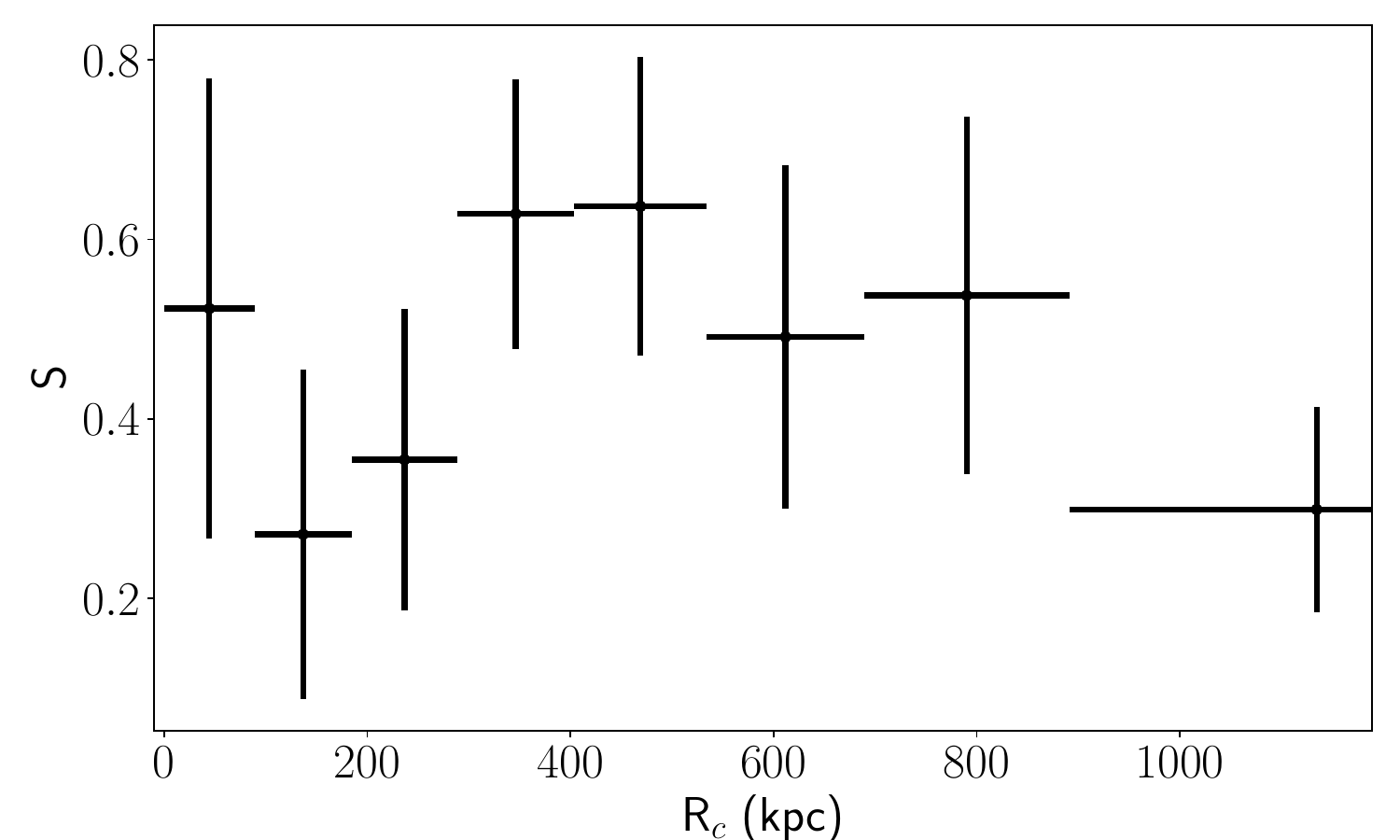}
\includegraphics[width=0.33\textwidth]{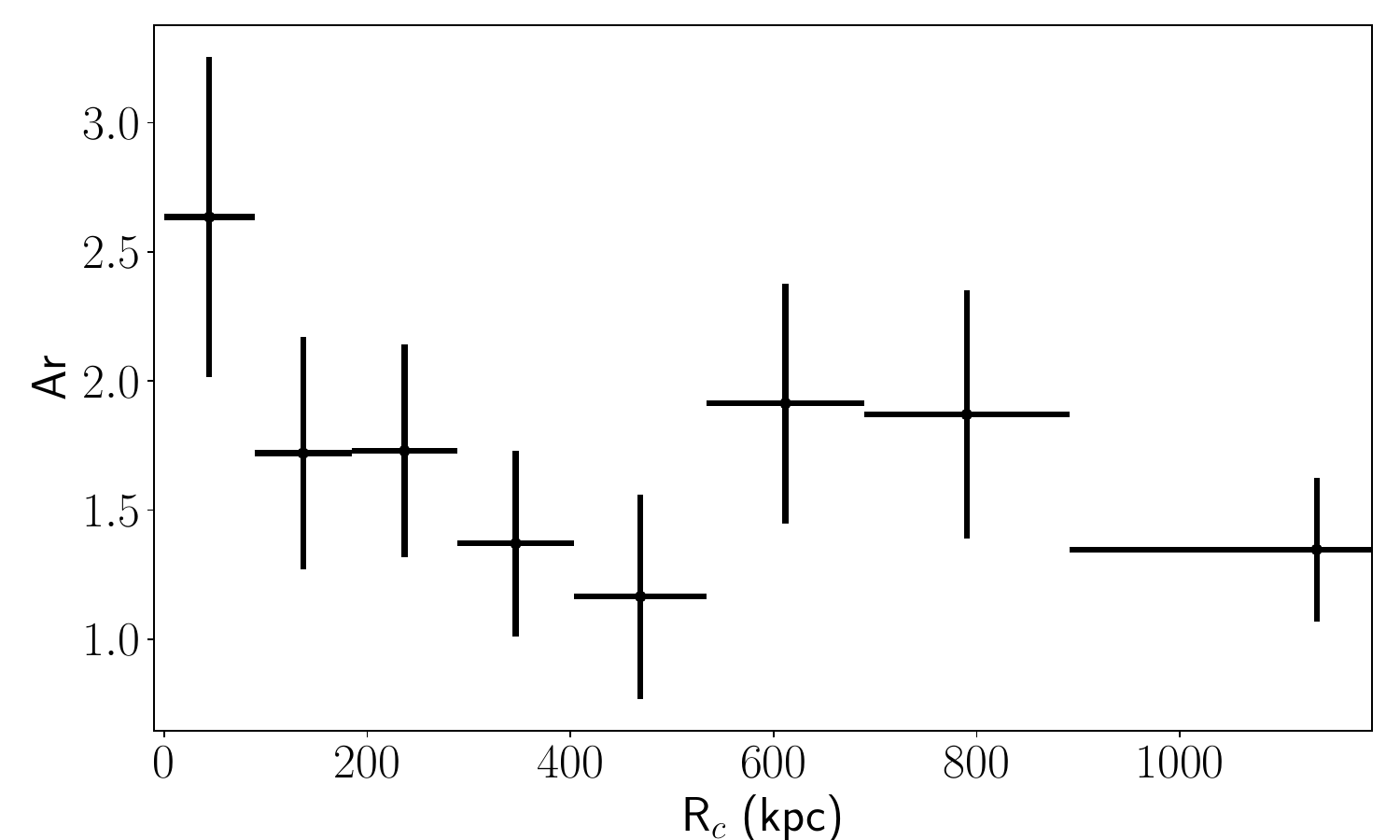} 
\includegraphics[width=0.33\textwidth]{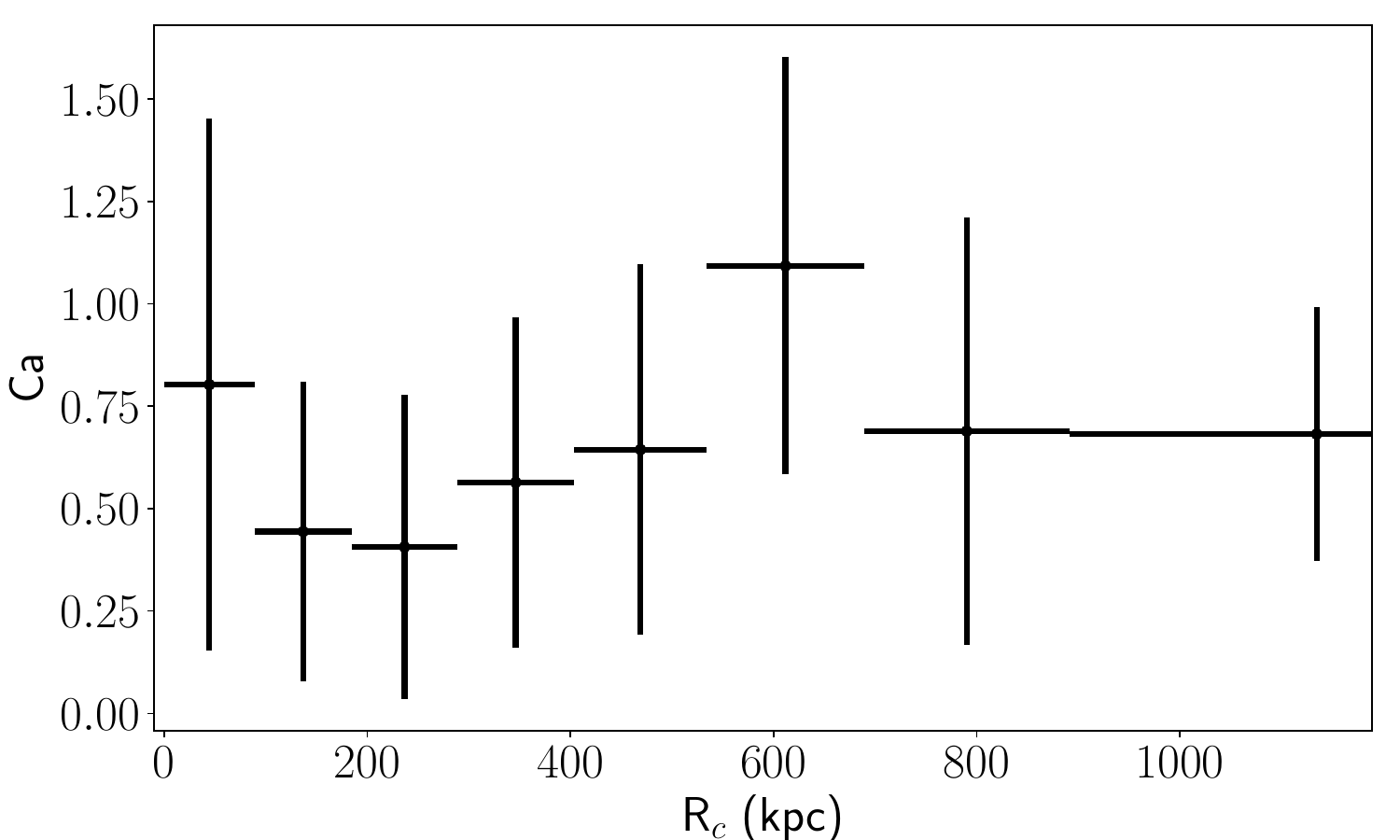}\\  
\includegraphics[width=0.33\textwidth]{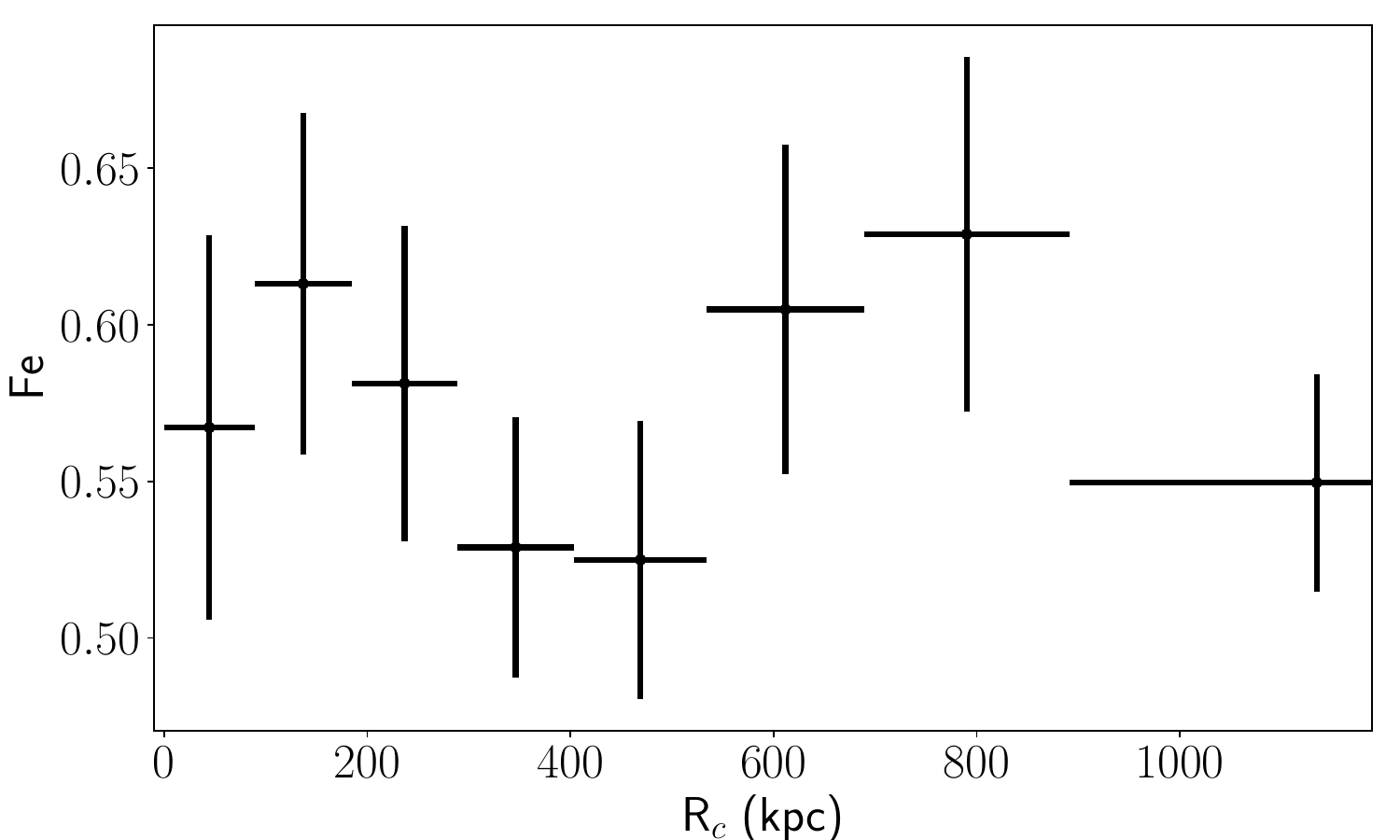}  
\caption{
A3266 abundance profiles were obtained from the local fits.
} \label{fig_abund_all} 
\end{figure*}

The best-fit elemental abundances obtained per region with respect to the Solar values are shown in Figure~\ref{fig_abund_all}.
For comparison, Figure~\ref{fig_abund_together} shows all elements in a combined plot. 
This is the first study of elemental abundances using X-ray observations within the A3266 cluster. 
The abundance uncertainties are somewhat large, most likely because the emission lines show low strength due to the high temperature, and therefore high ionization state, of the ICM \citep[for comparison with low-temperature systems such as Virgo and Centaurus see ][]{gat23b,gat23d}.
O, Mg, and Si derived abundances display positive gradients for distances $<600$~kpc before they become flatter. 
The Ca abundance displays a flat profile for lower distances.
All elements show a similar abundance in the $600-800$~kpc range, similar to the trend observed in the temperature.

In general, the abundance profiles display similar discontinuities to those obtained for the temperature (see Figure ~\ref{fig_kt_sigma})
Interestingly, the Fe profile does not display a large dispersion, a trend also identified in {\it Chandra} observations \citep{hen02}.
The high metallicity at large distances may indicate the mixing of metal-rich gas in the outskirts, most likely due to complex merger activities \citep{sun02,urd19,wal22}.
A similar trend was identified in Ophiuchus, another system with merging features \citep{gat23e}.
We have neither found any drop in the metallicity for the inner region of the cluster \citep[see, for example, the Centaurus analysis in ][]{gat23d}.
However, the spatial scale of the innermost region analyzed is much larger than the physical radius for which a drop in Fe abundance has been identified in other sources.
Finally, we have not found a correlation between the abundance and the velocity profiles. 
In order to better understand the dynamics and kinematics within such a system, future comparisons with detailed hydrodynamical simulations, including cluster merging events, are crucial.

The abundance determination using EPIC-pn implies limitations and caveats.
Despite the effort to improve the calibration \citep{dep07,rea14,sch15a}, cross-calibration differences between EPIC instruments have been reported.
Low-temperature systems (i.e., $kT\sim 0.7$~keV) tend to display a strong degeneracy between the emission measure and the metallicities \citep{mer22}.
The so-called Fe-bias occurs when a single-temperature plasma model is used for a multi-temperature ICM condition \citep{buo98,mer18,gas21}. 
However, we found that the best-fit abundances obtained with the {\tt vapec} model agree with those obtained with the {\tt vlognorm} model within $1\sigma$, although uncertainties tend to be lower for the multi-temperature model.
Other limitations may arise from systematic uncertainties associated with the astrophysical or instrumental background.
In that sense, the global-local fit procedure followed in this work, and fitting simultaneously pn and MOS instruments, is considered the most robust one \citep{mer15,fuk22}.

\subsection{ICM chemical enrichment from SN}\label{sec_snr}  
\begin{figure*}    
\centering
\includegraphics[width=0.33\textwidth]{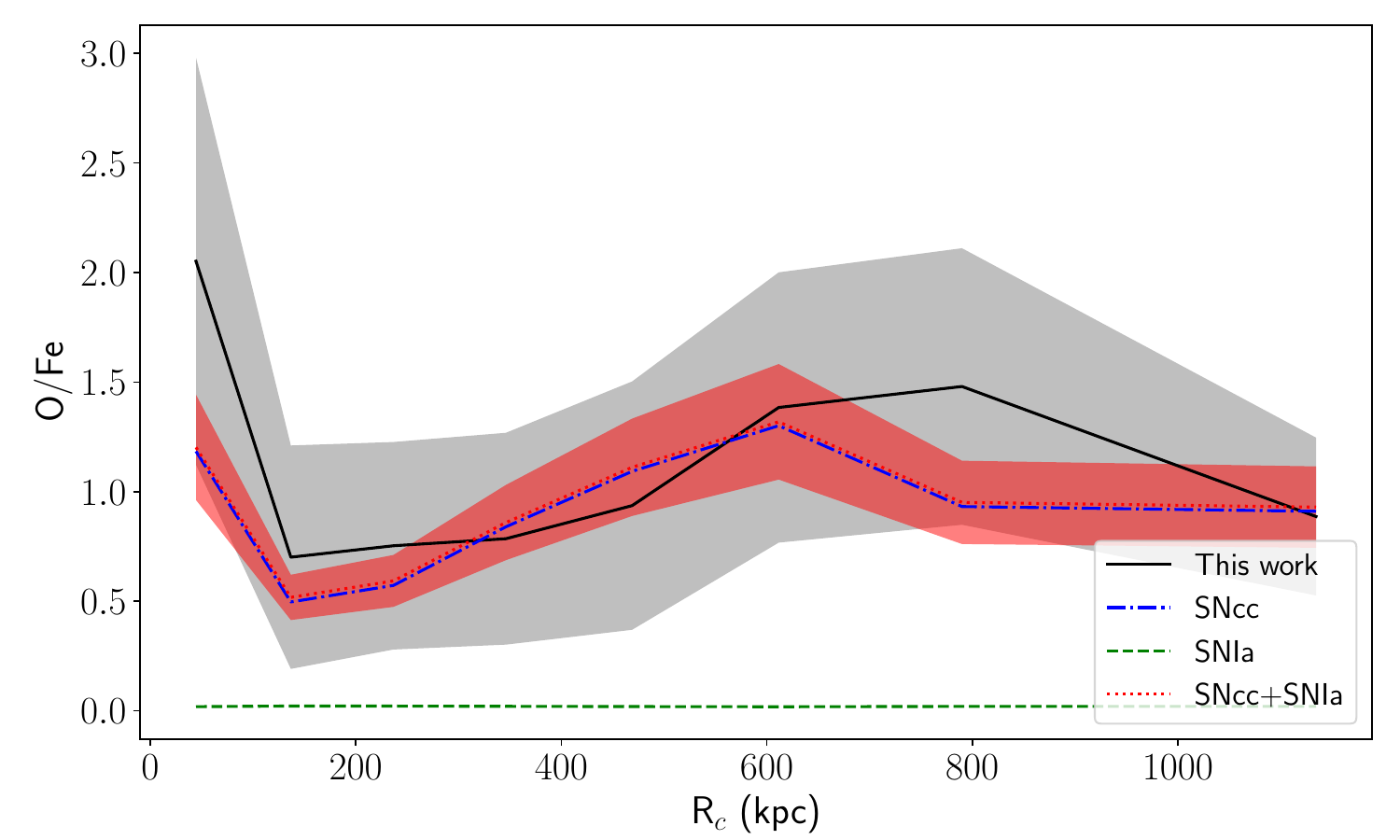}
\includegraphics[width=0.33\textwidth]{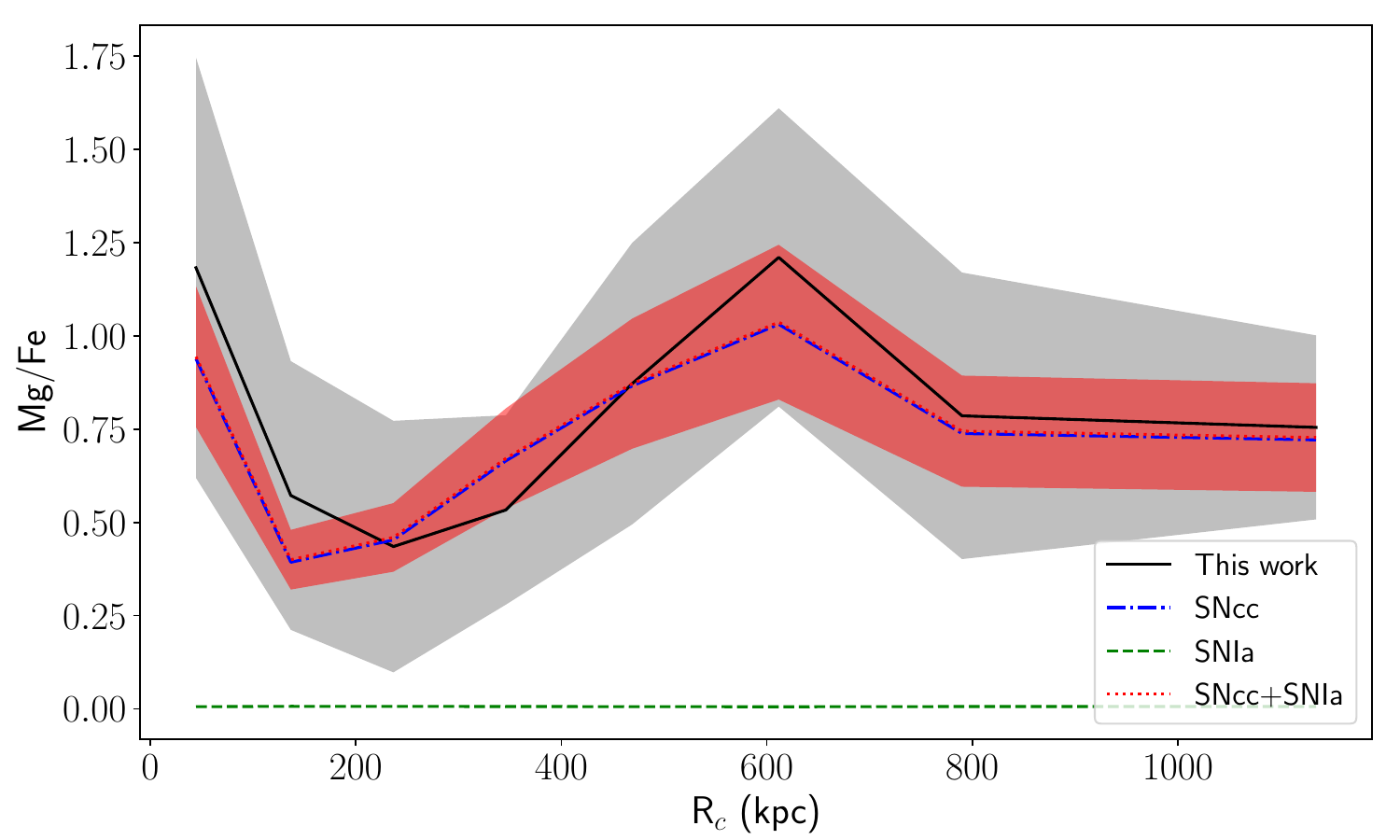} 
\includegraphics[width=0.33\textwidth]{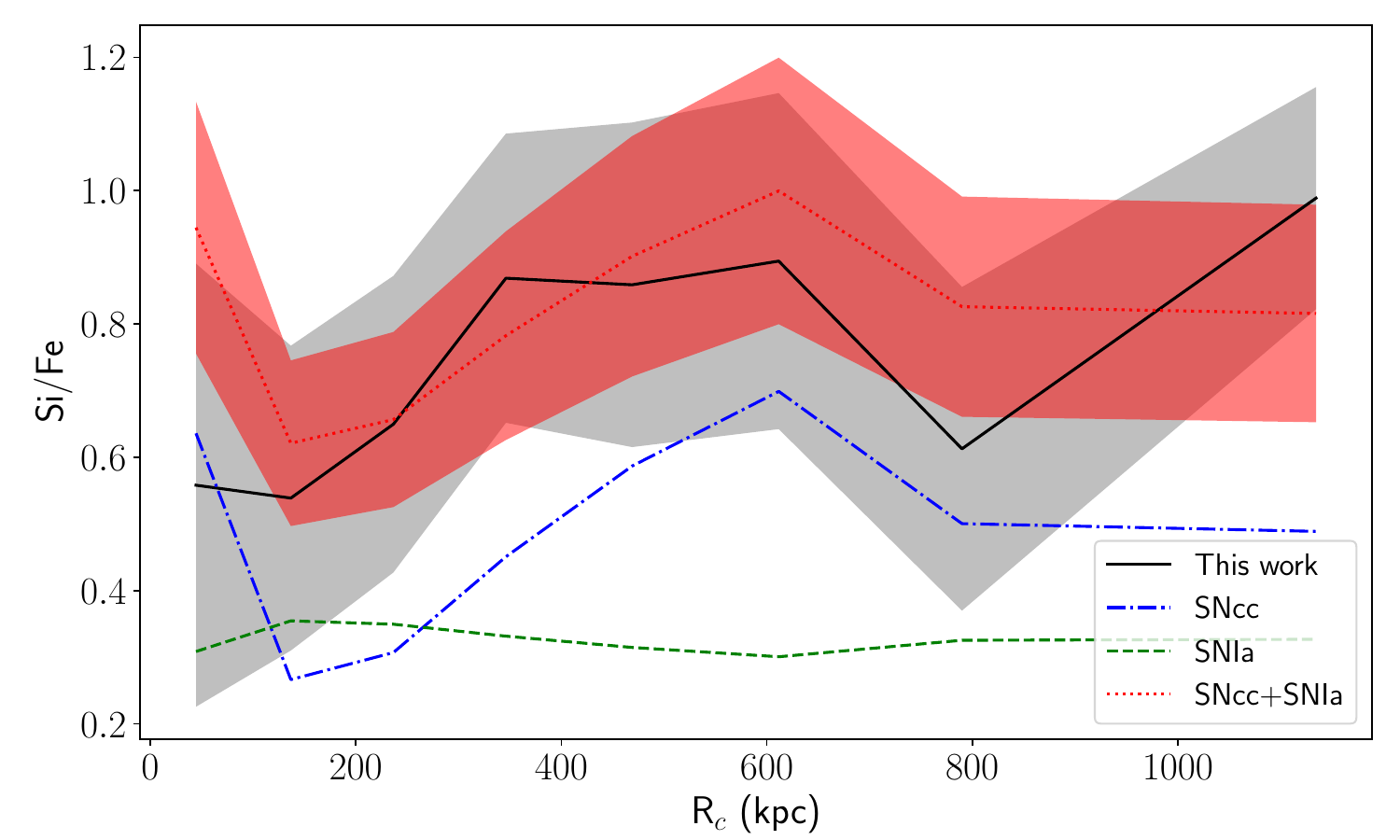} \\
\includegraphics[width=0.33\textwidth]{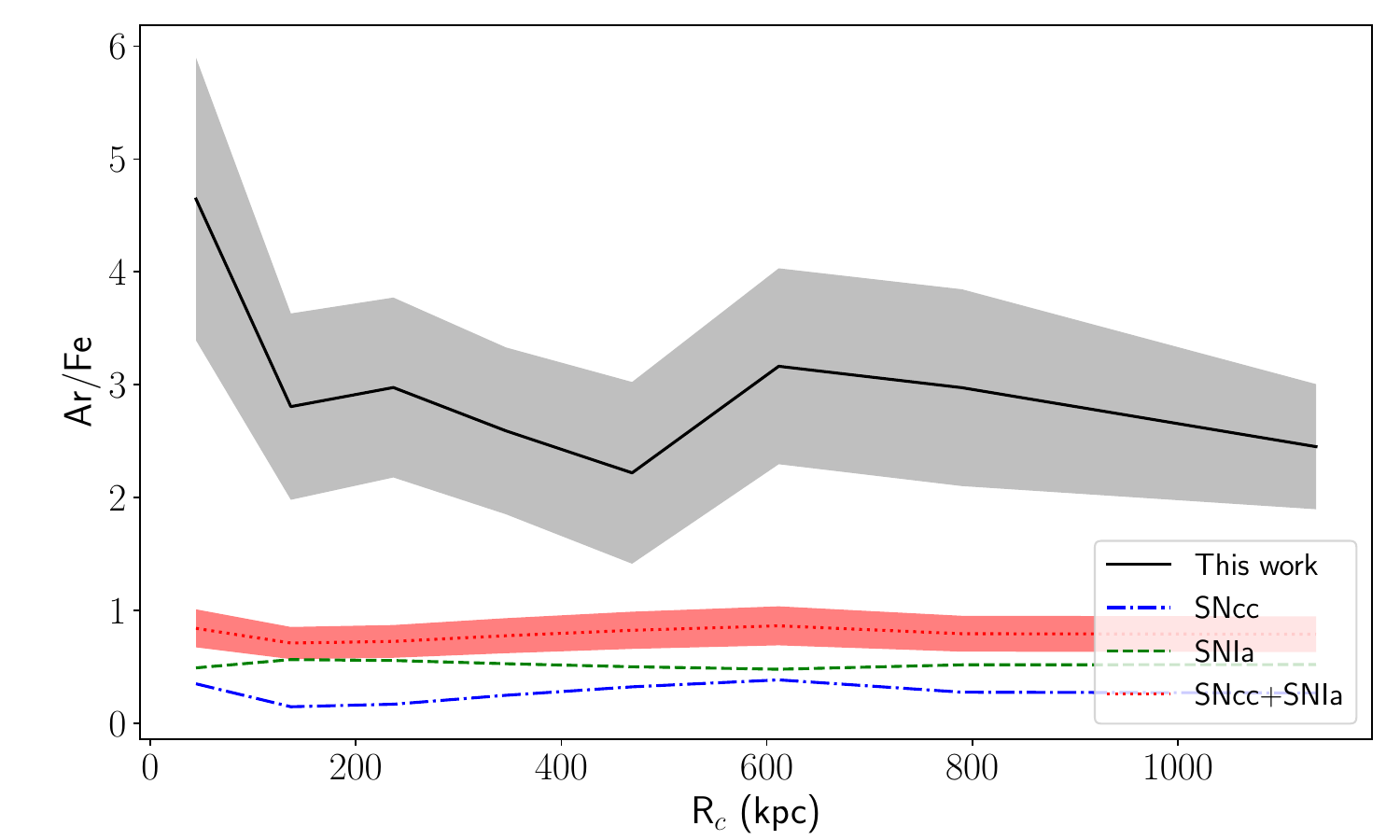}
\includegraphics[width=0.33\textwidth]{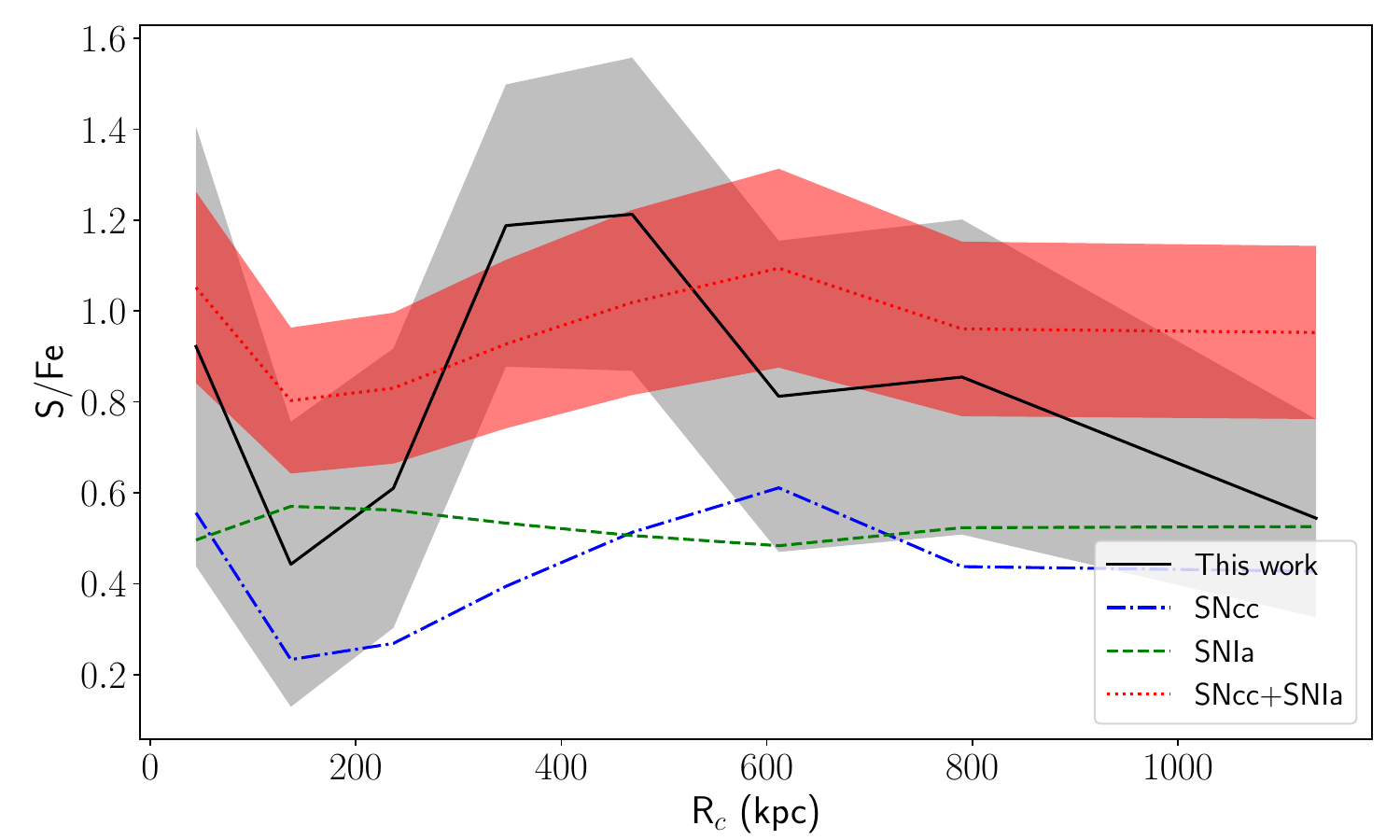} 
\includegraphics[width=0.33\textwidth]{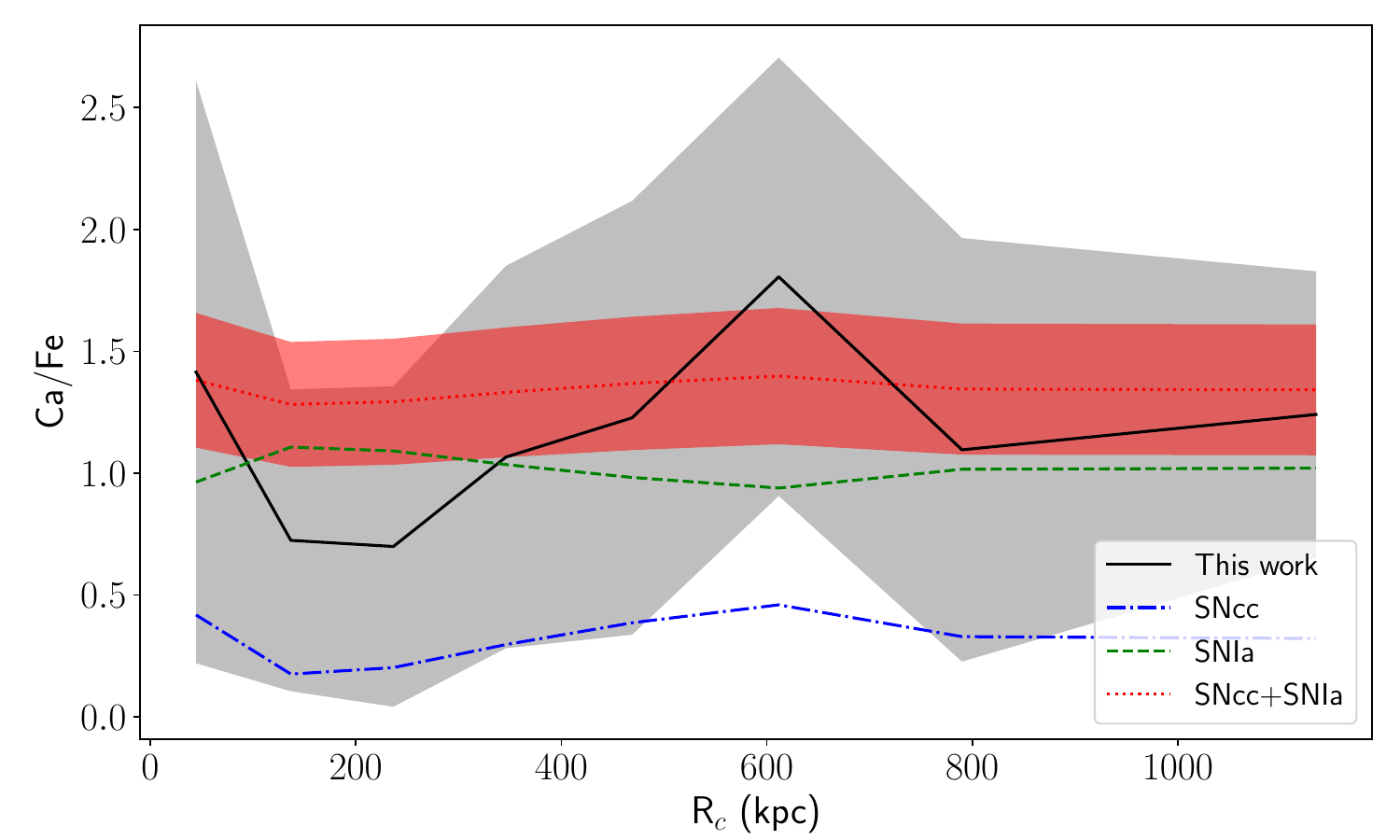} 
\caption{
Abundance ratio profiles, relative to Fe, obtained from the local fits. 
The gray shaded areas indicate the mean values and the 1$\sigma$ errors. 
The SNcc (blue line) and SNIa (green line) contribution to the total SN ratio (red shaded area) from the best-fit model are included (see Section~\ref{sec_snr}).
 } \label{fig_ratios} 
\end{figure*}

We computed X/Fe ratio profiles for O, Mg, Si, Ar, S, and Ca (gray shaded regions in Figure~\ref{fig_ratios}).  
Considering the uncertainties, we found that O/Fe, Mg/Fe, S/Fe, and Ca/Fe abundance ratios are close to the proto-solar values for all distances.
On the other hand, the Si/Fe tends to be lower, and the Ar/Fe is higher ($>2$ Solar) for all distances.
\citet{mer15} indicated that Ar abundances most likely suffer from instrumental systematics when comparing EPIC-pn with MOS spectra. 

We used the {\tt abundfit.py} python code developed by \citet{mer16b} to model the contribution from different SN yield models to the abundance ratio profiles obtained.
The code fits a combination of multiple progenitor yield models to compute the relative contribution better fitting a given set of ICM abundances.
For the SNcc and SNIa yields, we have included multiple models listed in Table~\ref{tab_snr_models}.
The SNcc yields were integrated with Salpeter IMF, while for SNIa, we considered models that included a variety of explosion mechanisms.
Such models have been used in previous enrichment studies \citep{mer17,sim19,mer20,gat23b,gat23d,gat23e}.
However, it is worth mentioning that further effort in improving theoretical models of supernova nucleosynthesis is essential to reduce uncertainties in the yield calculations (e.g., including neutrino physics). 

\begin{figure}    
\centering
\includegraphics[width=0.48\textwidth]{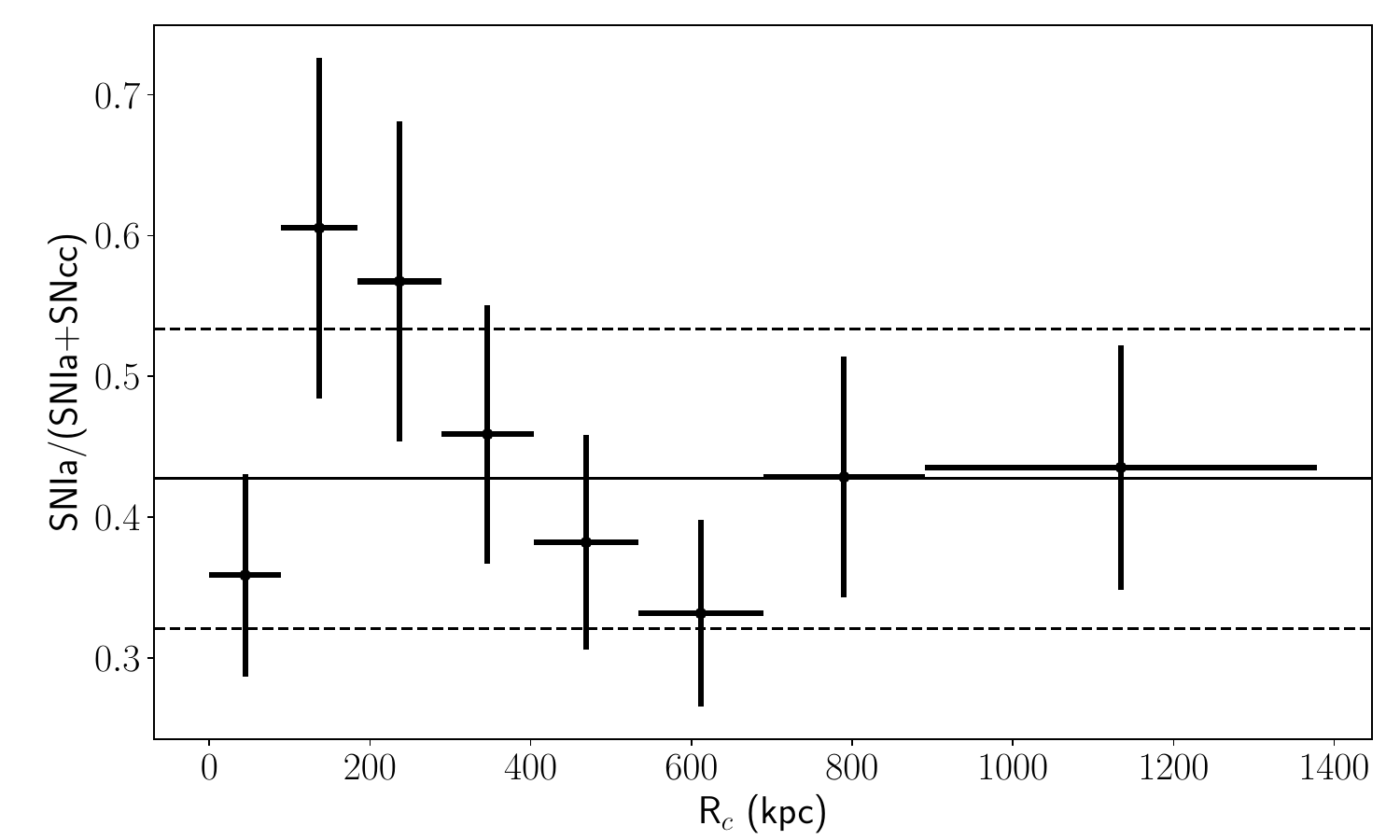} 
\caption{
SNIa contribution to the total chemical enrichment as a function of the distance. 
The model includes  O, Mg, Si, S, Ar, Ca, and Fe abundances. 
The black line represents the 1$\sigma$ confidence interval of constant fit.
} \label{fig_snia_distribution} 
\end{figure}

We use the same model along the radii to represent the profile, allowing us to determine the SNe ratio profile at all measured distances.
Then, we determined the best linear combination of SNIa and SNcc models that better fit the abundance ratio profiles by minimizing the sum of their $\chi^{2}$ values in quadrature.
The best-fit model ($\chi^{2}/d.o.f.=50.1/48=1.04$) corresponds to \citet{nom06} with initial metallicity Z$=0.01$ for the SNcc and the dynamically-driven double-degenerate double-detonation by \citet{she18} with parameters $M10,3070, Z=0.01$ for the SNIa contribution.
Figure~\ref{fig_ratios} shows the SNIa contribution (green line) and the SNcc contribution (blue line) to the total SNe ratio (red shaded region).
We found that the SNcc contribution is more significant than the SNIa contribution for almost all radii.
The contribution from SNIa to the total enrichment by this set of models is shown in Figure~\ref{fig_snia_distribution}.
A linear fit gives almost zero slope (2.69$\times$10$^{-5}$ with $\chi^{2}=1.14$) and a constant value of $0.42\pm 0.05$, verifying that the SNIa contribution to the total SNe tends to be constant.
Such a consistent radial profile may suggest that the SNIcc and SNIa components have occurred at similar epochs. 

These results support an early ICM enrichment scenario, with most of the metals present being produced prior to clustering, in agreement with our previous results for Virgo \citep{wer06,sim10b,mil11,gat23b}, Centaurus \citep{san16b,mer17,gat23d} and Ophiuchus \citep{gat23e}.
Compared with those previous works, our abundance ratio shows less scattering, most likely due to the inclusion of MOS spectra in the spectral fit.
As seen in Figure~\ref{fig_ratios}, the SNe model reproduces the observed pattern abundance for almost all elements except for Ar.
It is essential to mention that different contributions of the same SNIa and SNcc models also lead to good fits when compared with the measured abundance ratios (i.e., $\chi^{2}<1.2$).
Therefore, such models can only be dismissed partially.

The observed enrichment pattern indicates that the merging progenitors had comparable chemical histories, with most ICM metals produced before the cluster assembly. 
Furthermore, the absence of distinct metal-rich cores breaking apart during the merger implies significant mixing of the ICM, likely driven by merger-induced turbulence and shocks. 
Such behavior contrasts with non-merging clusters, where more pronounced metallicity gradients are often observed. 
These findings emphasize the role of mergers in shaping the chemical properties of the ICM and highlight the need for simulations that incorporate merger dynamics better to understand the interplay between cluster evolution and chemical enrichment.

 \begin{table}
\caption{SNIa and SNcc yields included in the modeling of the X/Fe profiles. }\label{tab_snr_models}
\begin{center}
\begin{tabular}{ll}
\hline
\hline
SNIa & SNcc     \\  
\hline  
\citet{iwa99} & \citet{chi04}\\
\citet{bad06} & \citet{nom06}  \\
\citet{mae10} & \citet{rom10} \\
\citet{sei13} & \citet{nom13} \\
\citet{fin14} & \citet{heg10} \\
\citet{ohl14} & \citet{suk16}\\
\citet{leu18} &  \\
\citet{sim12} &  \\
\citet{sei16} & \\
\citet{mar15} &  \\
\citet{kro15} &  \\
\citet{wal11} &  \\
\citet{sim10a} & \\
\citet{pak10} &\\
\citet{pak12} & \\
\citet{kro13} & \\
\citet{she18} & \\
\hline 
\end{tabular}
\end{center} 
\end{table}

\section{Conclusions and summary}\label{sec_con} 
We have analyzed {\it XMM-Newton} EPIC observations of the A3266 galaxy cluster to study the radial profiles of physical parameters such as temperature and the O, Mg, Si, S, Ar, Ca, and Fe distribution in the ICM.
This is the first analysis of such profiles for A3266.
This work expands upon the outcomes of \citep{gat24a} by including the soft energy band $<4$~keV and MOS spectra.
Our main findings and conclusions are:
\begin{itemize}
\item We model the ICM emission using a single temperature model. 
From a statistical point of view, there is no improvement in the fit when including a more complex model, most likely because of the high temperature of the system. 
\item We found velocities in good agreement with \citep{gat24a}. 
Moreover, including MOS spectra leads to lower uncertainties $\Delta v\sim 80$~km/s.
There is no clear trend between the velocities and other physical parameters such as temperature or abundance.
\item The temperature show discontinuities at $\sim 350$~kpc and $\sim 800$~kpc.
They could be associated with the presence of subgroups.
\item We obtained abundance profiles for O, Mg, Si, S, Ar, Ca, and Fe. 
Most of the elements display the same discontinuities found for the temperature.
\item We model X/Fe abundance ratio profiles with a linear combination of SNcc and SNIa.
We found that the best-fit model corresponds to a dynamically-driven double-degenerate double-detonation for the SNcc contribution and an SNcc model with an initial metallicity of $Z=0.01$.
This model reproduces the abundance ratios, except for Ar.
\item We found that the SNIa ratio over the total cluster enrichment tends to be uniform.
Such uniformity in the SNIa percentage contribution supports an early enrichment of the ICM scenario, where most of the metals present were produced before clustering.
\end{itemize}

\begin{acknowledgements} 
This work was supported by the Deutsche Zentrum f\"ur Luft- und Raumfahrt (DLR) under the Verbundforschung programme (Messung von Schwapp-, Verschmelzungs- und R\"uckkopplungsgeschwindigkeiten in Galaxienhaufen.). 
This work is based on observations obtained with XMM-Newton, an ESA science mission with instruments and contributions directly funded by ESA Member States and NASA. 
This research was carried out on the High-Performance Computing resources of the Raven and Viper clusters at the Max Planck Computing and Data Facility (MPCDF) in Garching, operated by the Max Planck Society (MPG).  
\end{acknowledgements}

\bibliographystyle{aa}

\end{document}